\documentclass[floatfix,aps,prx,reprint,groupedaddress,twocolumn]{revtex4-2}
\usepackage{color}
\usepackage{graphicx}
\usepackage{amsmath,amssymb,wasysym}
\begin{document}

\title{Canonical Simulation Methodology to Extract Phase Boundaries of Liquid Crystalline Polymer Mixtures}
\author{William S. Fall}
%\email{william.fall@sheffield.ac.uk}
\affiliation{Department of Physics and Astronomy, University of Sheffield, Sheffield, S3 7RH, United Kingdom.}

\author{Hima Bindu Kolli}
%\email{h.bindukolli@sheffield.ac.uk}
\affiliation{Department of Physics and Astronomy, University of Sheffield, Sheffield, S3 7RH, United Kingdom.}

\author{Biswaroop Mukherjee}
%\email{b.mukherjee@sheffield.ac.uk}
\affiliation{Department of Physics and Astronomy, University of Sheffield, Sheffield, S3 7RH, United Kingdom.}

\author{Buddhapriya Chakrabarti}
\email{b.chakrabarti@sheffield.ac.uk}
\affiliation{Department of Physics and Astronomy, University of Sheffield, Sheffield, S3 7RH, United Kingdom.}

\date{\today}

\begin{abstract}
We report a novel multi-scale simulation methodology to quantitatively predict the thermodynamic behaviour of polymer mixtures, that exhibit phases with broken orientational symmetry. Our system consists of a binary mixture of oligomers and rod-like mesogens. Using coarse-grained molecular dynamics (CGMD) simulations we infer the topology of the temperature-dependent free energy landscape from the probability distributions of excess volume fraction of the components. The mixture exhibits nematic and smectic phases as a function of two temperature scales, the nematic-isotropic temperature $T_{NI}$ and the $T_c$, the transition that governs the polymer demixing. Using a mean-field free energy of polymer-dispersed liquid crystals (PDLCs), with suitably chosen parameter values, we construct a mean-field phase diagram that semi-quantitatively match those obtained from CGMD simulations. Our results are applicable to mixtures of synthetic and biological macromolecules that undergo phase separation and are orientable, thereby giving rise to the liquid crystalline phases. 
\end{abstract}

\maketitle

\section{\label{S1:Intro}Introduction}

Complex mixtures of solute and solvent molecules are widespread, encompassing subjects ranging from physics and chemistry to materials science and even biology. These materials organise on a mesoscopic length scale, which lies between the smaller microscopic and larger macroscopic length scales and are inherently soft \cite{nagel2017experimental,van2018grand,evans2019simple}. This softness arises from relatively weak interactions ($\sim k_B T$) between molecular constituents and as such thermal fluctuations play a major role in deciding both their structural and dynamical behaviour. Therefore both entropic and enthalpic effects are important in determining their phase behaviour. The solute molecules can attract or repel each other and their relative strengths can be manipulated by changing the temperature or composition, which results in a series of different ordered self-assembled structures.

On going from spherically symmetric to anisotropic molecules an even richer phase behaviour \cite{de1993physics} is observed, not only controlled by entropy and enthalpy but also directional interactions between the anisotropic components. The simplest phase behaviour arises in polymer solutions, where the mixed state is stabilised by the entropy of mixing at higher temperatures. Upon lowering the temperature the enthalpic effects take over and below the bulk melting temperature $T_{c}$, it is energetically more favourable for the system to phase separate and exist as a mixture of polymer rich and solvent rich regions. In the reverse scenario, cooling polymer melts results in the appearance of a semi-crystalline polymeric glass phase, in which polymer chains are packed parallel to each other forming lamellar regions which coexist with amorphous regions with an imperfect packing \cite{de1993physics,evans2019simple,cingil2017illuminating,iwaura2006molecular,nagel2017experimental,van2018grand}. 

Polymer dispersed liquid crystals (PDLCs) are one such example and are an important class of materials with applications ranging from novel bulk phenomena in electro-optic devices \cite{bronnikov2013polymer} to very rich and unique surface phenomena like tunable surface roughness \cite{liu2015reverse} and electric field driven meso-patterning on soft surfaces \cite{roy2019electrodynamic,dhara2018transition}. These soft materials can be termed multi-responsive as they can be controlled by electro-magnetic fields, the presence of interfaces or substrates and temperature or concentration-gradients \textit{etc}. While there is a lot of literature available on the synthesis and application of these novel materials, a fundamental understanding of the thermodynamics and kinetics of phase transformations in these complex mixtures is still missing.  

\begin{figure}[htb]
    \centering
    \includegraphics[width=\columnwidth]{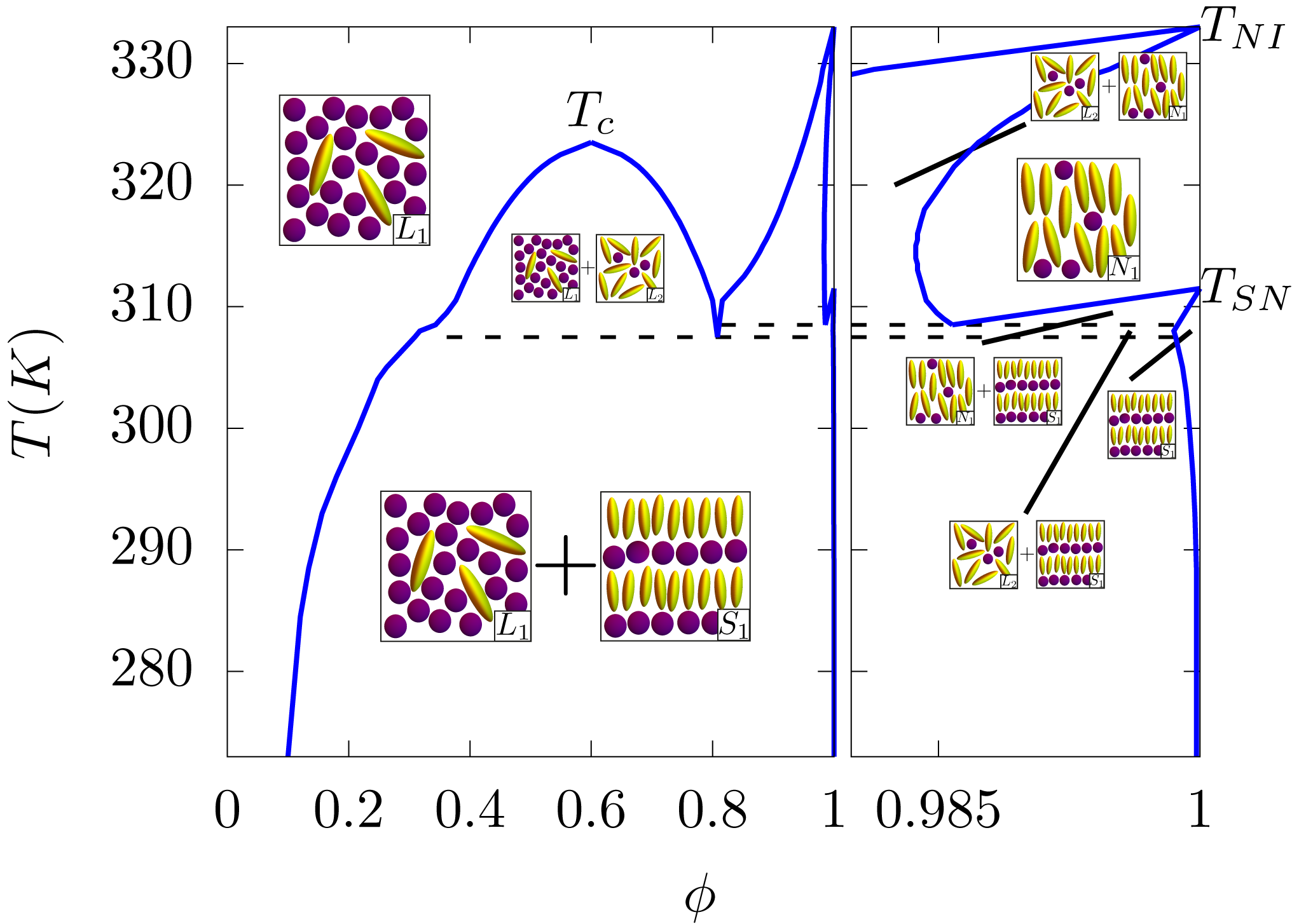}
    \caption{\label{fig:figure1} A typical ``teapot'' phase diagram for a mixture of longer flexible polymers and shorter rod-like smectic-A mesogens, reproduced from \cite{kyu1996phase}, where $\phi$ indicates the LC volume fraction and $T$ temperature. The inset panels illustrate the phases that can exist in each of the respective regions demarcated by the (blue) phase boundaries; the flexible polymers and rod-like mesogens are represented by (purple) dots and (yellow) oblate spheres respectively. The four different phases include: mesogen-poor liquid, mesogen-rich liquid, nematic and smectic as denoted by $L_{1}$, $L_{2}$, $N_{1}$ and $S_{1}$ respectively. Dashed lines mark the triple points, $T_{c}$ the (continuous) transition from single-phase to two-phase liquid, $T_{SN}$ the (first-order) transition from smectic to nematic and $T_{NI}$ the (first-order) transition from nematic to isotropic liquid. Parameters used: $T_{NI}=333K$, $\alpha=0.851$, $r_{2}/r_{1}=2.25$ and $\chi(T)=-1+772/T$.}
    \label{fig:fig_1}
\end{figure}

Interestingly, a rich phase diagram is also observed for complex mixtures with only repulsive interactions. The fundamental reason behind this phase behaviour was identified fairly early, in mixtures of rod-like hard particles and non-adsorbing polymers, from density functional theory (DFT) calculations \cite{lekkerkerker1992phase,lekkerkerker1994phase}. It is due to an effective depletion attraction between the rod-like particles when they are at small separation and thus, even in systems with only excluded volume interactions, one observes three distinct phases: of which two are isotropic ($L_{1}+L_{2}$) (one polymer-rich and the other mesogen-rich) and one mesogen-rich nematic $N_{1}$. Simple mean-field models of mixtures of polymers and liquid crystals has, however, considered both entropic and enthalpic effects \cite{mcmillan1971simple,chiu1998phase,matsuyama2002non}. The phase boundary between the one-phase and the two-phase regions of these mixtures in the temperature-order parameter plane (shown in Figure \ref{fig:fig_1} by the blue line) is commonly referred to as having a ``teapot" topology and is characterised by a number of special points. The primary order parameter, $\phi$, is the difference between the local densities of the two components, the polymers and the liquid crystals. The ``top" of the teapot is the critical point and its "lid" is coexistence of the polymer-rich and the mesogen-rich isotropic phases (see Figure \ref{fig:fig_1}). In this region, 
the order parameter $\phi$, which is of the Ising universality class, grows continuously from zero as one cools the system below $T_c$. At order parameter values close to unity, the system consists primarily of the mesogens. For a purely mesogenic phase ($\phi = 1$), as one cools the system the nematic order parameter discontinuously jumps to a non-zero value, at the isotropic-nematic transition temperature, $T_{NI}$ which forms one ``spout" of the teapot. At this point the rotational invariance of the configurations are broken and the mesogens spontaneously order along a common director. This order parameter belongs to a different universality and the phases formed by the mixture of polymers and liquid crystals allow a novel interplay between order parameters of different symmetries which affects both the thermodynamics and kinetics of these complex mixtures. In some situations, cooling the system further results in the sudden appearance of a non-zero smectic order at $T_{SN}$, a thermodynamic state characterised by broken orientational symmetry and a one dimensional positional ordering. The phases coexisting in this region are pictorially shown in  Figure \ref{fig:fig_1}.
The relative positions of these special points in the temperature-composition plane, which again are functions of the strengths of the microscopic interactions, shape the phase diagram. The four different phases appearing here are : mesogen-poor liquid, mesogen-rich liquid, nematic and smectic as denoted by $L_{1}$, $L_{2}$, $N_{1}$ and $S_{1}$ respectively and their coexistence regions are also shown in Figure \ref{fig:fig_1}. For a more detailed discussion on how the shape of the phase boundaries is affected by the parameter values please refer to Section 4 of Appendix C. 

Recently, the nematic ordering of semi-flexible macromolecules, in implicit solvents, have been studied in the limit where the contour length, $L$, is much greater than its persistence length $l_{p}$ using large-scale molecular dynamics simulations. Owing to large director fluctuations, the effective tube radius within which each macro-molecule is confined is much greater than what should be expected from the length scale arising from average density \cite{egorov2016anomalous}. These director fluctuations modify the phase diagram one computes from density functional theories. In material systems both entropic and enthalpic interactions decide the phase behaviour of complex mixtures and an interplay between nematic order and phase separation has been recently studied for polymeric chains in implicit solvents of varying quality \cite{midya2019phase}. The stiffer chains showed a single transition from isotropic to nematic, while the softer chains also exhibited a demixing between isotropic fluids, one polymer-rich and the other mesogen-rich \cite{midya2019phase}. 

Phase diagrams are central to the understanding of material properties as the regions of thermodynamic stability of materials are encoded in them. Calculating phase diagrams from molecular simulations however is a task which is far from trivial \cite{frenkel2001understanding}. One of the most prominent methods is the Gibbs ensemble technique \cite{panagiotopoulos1987direct} which is used for computing phase diagrams of liquid-vapour systems and for fluid mixtures, with the method of thermodynamic integration being another \cite{grochola2004constrained}. A number of recent publications have introduced a powerful method for estimating the whole phase diagram from a single molecular dynamics simulation by leveraging the multithermal-multibaric ensemble \cite{valsson2014variational,piaggi2019calculation,piaggi2019multithermal}. 

In this work, we develop a multi-scale simulation methodology to map out the phase diagram of a binary mixture of rod-like mesogens in an explicit solvent of oligomers via CGMD simulations and qualitatively match it to its theoretical counterpart from mean-field theory \cite{chiu1998phase}. By scanning the 
temperature-composition space via multiple CGMD simulations and by monitoring the resulting order parameter distributions we infer about the boundary between the locally stable and unstable regions. This maps out the phase boundaries and by appropriately tuning parameters appearing the mean-field theory we obtain a phase diagram which is qualitatively similar to the CGMD phase diagram. In principle, this method can be applied to a host of soft matter systems involving ordering fields competing
with phase separation like associating fluids like gels, gel-nematic mixtures, nematic-nematic mixtures etc. The remaining paper is organised into the following sections : Section II discusses our simulation methodologies, section III discusses the results of the molecular dynamics simulations, the global order parameters, the mean-field phase diagram is discussed next, along with how the phase boundaries inferred from the analyses of the partial free-energies obtained from the CGMD trajectories qualitatively agrees with the mean-field phase diagram upon a reasonable choice of parameter values. In Section IV, we conclude.

\section{\label{S2:Method}The Methodology}
\subsection{Mapping Phase Boundaries}
Our method, developed to extract phase boundaries from MD simulation trajectories, proceeds as follows. Simulations of a binary system, in this case a mixture of rod-like mesogens and oligomers, are performed for a series of initial starting compositions $\phi_{0}$, at high temperature and quenched to carefully chosen points in the $T-\phi$ plane. Details of the simulation results and coarse-grained model, including the parameters used, can be found in Section \ref{S3:Results} and Appendix \ref{Appendix:CGMD_Model} respectively. For a given volume fraction of the LC component $\phi_{0}$, the system will phase separate depending on its location in the underlying free energy landscape. In order to probe the topology of the underlying landscape, a new procedure has been devised.

From the resulting simulation library an estimate of the correlation length $\xi$, is first made, for a set of independent quenches at a given point and used to inform a specially devised binning procedure. Each trajectory is then binned into cubes such that the composition of the cubes may be evaluated, using a suitably defined continuum order parameter, to produce a histogram of the continuum order parameter distribution, $P(\phi; \phi_{0})$. The extracted distributions are then inverted to reveal a partial free energy $f(\phi; \phi_{0})=-k_{B}T\log P(\phi; \phi_{0})$ containing several minima, which consequently expose the topology of the free energy landscape and the approximate location of the phase boundaries.

\begin{figure}[htb]
\includegraphics[]{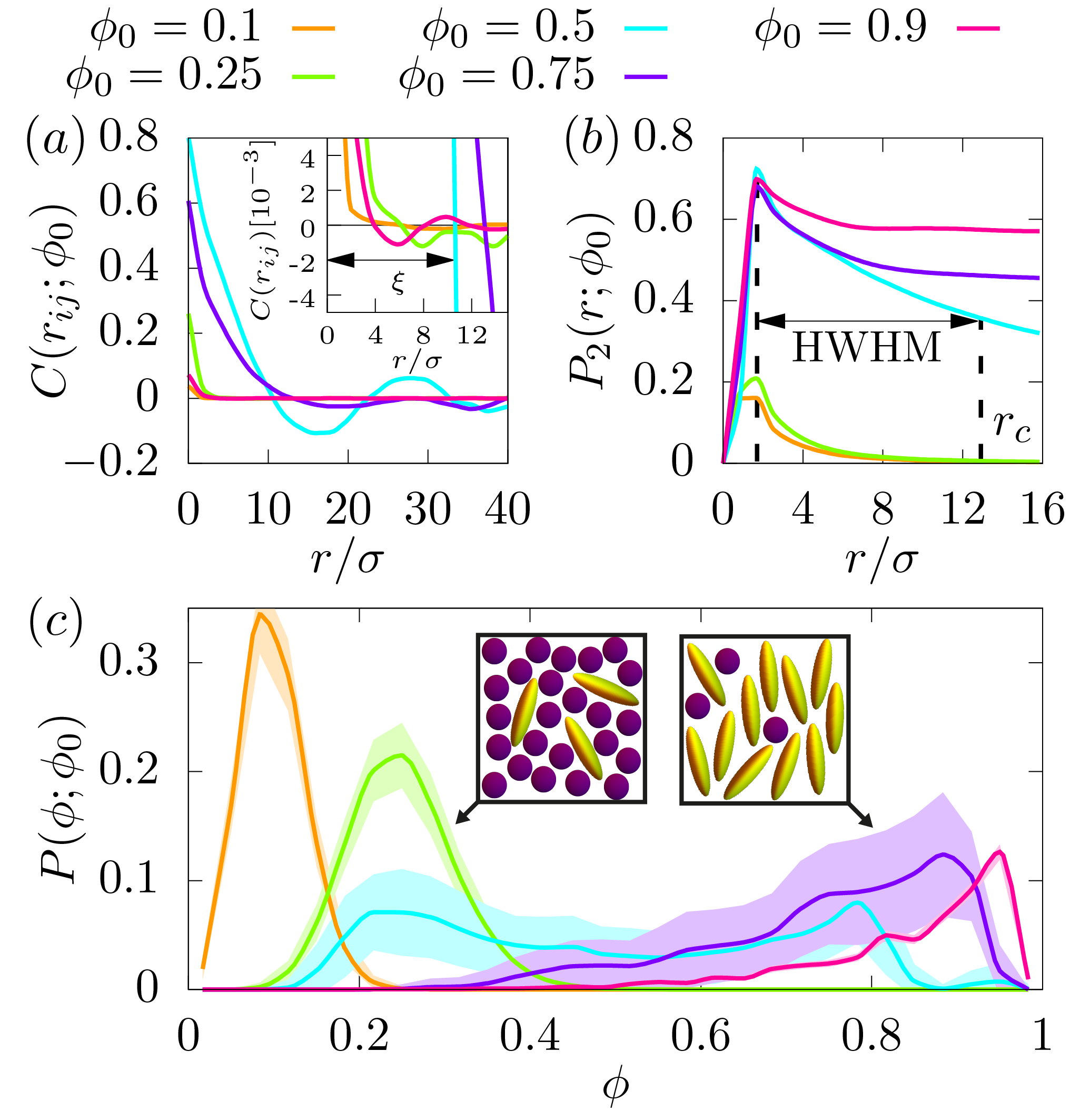}
\caption{\label{fig:figure2} Correlation functions, local nematic order parameter and continuum order parameter distributions from MD simulations at $T^{*}=5.1$. (a) Correlation function used to estimate the correlation length and bin size. The zoomed inset shows the zero-crossing points more clearly for all compositions and the correlation length $\xi$, for the $\phi_{0}=0.5$ composition, is indicated by the arrows as an example. (b) Local $P_{2}$ order as a function of the cutoff distance, the HWHM is indicated and used as the cutoff distance $r_{c}$ when assigning $P_{2}$ values to each rod.(c) The probability distribution as a function of the density, the cartoon panels indicate the mesogen-rich and mesogen-poor regions where the flexible polymers and rod-like mesogens are represented by (purple) dots and (yellow) oblate spheres respectively.}
\end{figure}

The first step of our numerical recipe is to determine the correlation length at each point under consideration in the $T-\phi$ plane. This is achieved by coarse-graining the order parameter field and effectively reducing it to a spin-1/2 Ising-like configuration. Each of the instantaneous simulation snapshots are binned into cubes of size $\approx (2\sigma)^{3}$, $\sigma$ is defined in Appendix \ref{Appendix:CGMD_Model} and the number of monomers of each species $n_{A}$ and $n_{B}$ inside are totalled. A state $\Psi=\pm1$ is then assigned to each cell such that $\Psi=1$ if $n_{A}>n_{B}$ and $\Psi=-1$ otherwise. The spatial correlation function is then calculated, 
\begin{eqnarray}
C(r_{ij})=(\Psi_{i} - \langle\Psi\rangle)(\Psi_{j} - \langle\Psi\rangle)
\label{e:correlation}
\end{eqnarray}
where $r_{ij}$ is the radial distance between the respective cubes and the angle brackets indicate averaging over a suitable long time period, in this case the last 20ns of all independent quenches are used. Figure \ref{fig:figure2} (a) depicts typical correlation functions calculated from MD simulations as quenched from $T^{*}=10.5$ to $T^{*}=5.1$ for all compositions considered in this work. The zero-crossing point for each composition indicates the correlation length $\xi$ as indicated explicitly for the $\phi_{0}=0.5$ composition in the figure. The correlation length for each of the simulations considered then serves as a customised estimate for the bin size used in the subsequent binning procedure to determine the continuum order parameter distribution $P(\phi; \phi_{0})$.

In the second step, the order parameter distribution is determined by re-binning the simulation cell into cubes with dimensions $\approx\xi^{3}$, this is illustrated in Figure \ref{fig:figure7} (a). The number of monomers of each species inside each bin are counted and a value assigned, using the order parameter of an arbitrary bin $i$, which is defined as
\begin{eqnarray}
\phi_{i}=\frac{1}{2}\Bigg( \frac{n_{A}^{i}-n_{B}^{i}}{n_{A}^{i}+n_{B}^{i}} + 1\Bigg)
\end{eqnarray}
In this case however, the continuum order parameter $\phi_{i}$, is bounded between zero and unity and is the continuum definition of the order parameter $\Psi$, defined above. The probability distribution $P(\phi;\phi_{0})$ may be found by averaging this process over the last 20ns of each independent quench and producing a histogram of the bin values. Figure \ref{fig:figure2} (c) shows typical probability distributions which reveal a distinct splitting of the simulation cell to its bracketing densities, revealing a series of mesogen-rich and mesogen-poor regions as indicated by the inset cartoon panels.

In the final step the order parameter distributions are inverted to reveal the topology of the free energy landscape though a partial free energy $f(\phi;\phi_{0})=-k_{B}T\log P(\phi;\phi_{0})$ at each composition, $\phi_{0}$. Figure \ref{fig:figure3} (e) shows an example inversion from MD simulations from the different compositions at $T^{*}=5.1$. A series of minima are present indicating the system is splitting to lower its free energy. In Appendix \ref{Appendix:Methods_Rationalise} this is rationalised using a conserved order parameter dynamics model which links the starting composition to the nature of the probability distribution of the order parameter at long times, which in turn is linked to the topology of the underlying free energy landscape. This leads to an important rule which may be used to understand the resulting partial free energy profiles. Simulations that converge onto their starting compositions $\phi_{0}$ with a single minimum are initiated from region of positive curvature, or $f^{\prime \prime} (\phi;\phi_{0}) > 0$ and those with that split into two or more successive minima are initiated from a region of negative curvature $f^{\prime \prime} (\phi;\phi_{0}) < 0$ and spontaneously phase separate. In Section \ref{S3:Results_MFTvsMD} this rule is employed to map out the approximate location of the phase boundaries from our CGMD simulations. 

\subsection{Characterising Phase Boundaries}

The second half of our numerical recipe is concerned with identifying which liquid or liquid crystalline phase each of the respective minimums, in the partial free energy profiles, correspond to. Depending on whether or not the system splits or converges onto its equilibrium composition a local or global approach must be used respectively, in order to determine the extent of orientational ordering of the rod-like mesogens. In the scenario where the system splits between different ordered and disordered phases, a global approach cannot be used to determine the type of LC phase since it would mask the locally ordered nematic regions. Instead the local nematic order $P_{2}(r)$ of each rod-like molecule is probed as a function of the cutoff distance $r_{c}$,\cite{mukherjee2012derivation,cinacchi2009diffusivity},
\begin{eqnarray}
\scalebox{1.05}[1]{$P_{2}(r)=\Bigg\langle \frac{\Sigma^{N-1}_{i=1}\Sigma^{N}_{j=i+1}\delta(r-|\mathbf{r}_{j}-\mathbf{r}_{i}|)P_{2}(\hat{\textbf{u}_{i}}(\mathbf{r}_{i})\cdot\hat{\textbf{u}_{j}}(\mathbf{r}_{j}))}{\Sigma^{N-1}_{i=1}\Sigma^{N}_{j=i+1}\delta(r-|\mathbf{r}_{j}-\mathbf{r}_{i}|)} \Bigg\rangle$}
\end{eqnarray}
where $P_{2}$ is the second Legendre polynomial and $\hat{\textbf{u}_{i}}(\mathbf{r}_{i})$ the unit vector associated with the largest eigenvalue of the inertia tensor of particle $i$, with its centre of mass located at $\mathbf{r}_{i}$, see Appendix \ref{S2:Intro_Order} for methodological details. The angular brackets indicate statistical averaging over the last 20ns of each independent quench. Figure \ref{fig:figure2} (b) shows a series of $P_{2}$ curves as a function of the cutoff distance $r_{c}$ for each of the compositions considered at $T^{*}=5.1$. As a convention the half width half maximum (HWHM) is then used as the cutoff $r_{c}$, to assign a $P_{2}$ value to each rod-like mesogen in the system. For reference $P_{2}\sim1$ indicates perfect orientational order of the rods and $P_{2}\sim0$ a completely random orientation as illustrated in Figure \ref{fig:figure8}.

In order to then isolate the extent of nematic ordering within each of the distinct splitting regions, the $P_{2}$ ordering of the molecules is then coupled with the order parameter distributions $P(\phi;\phi_{0})$. This is achieved by isolating the bins at different points along the $P(\phi;\phi_{0})$ histogram and then averaging the local $P_{2}$ values of the molecules inside. In Figure \ref{fig:figure3} (f) points at different intervals along the $f(\phi;\phi_{0})$ profiles have been coloured from blue (isotropic) to red (anisotropic) according to their local $P_{2}$ values and consequently numerous different liquid and liquid crystalline phases are revealed. We note that it is possible for a bin to have a non-zero continuum order parameter $\phi_{0}$ and return a null local nematic order parameter $P_{2}$ value. In this situation there are no rods in the system with a centre of mass (COM) that lie inside the bin and thus the $P_{2}$ values cannot be averaged. The continuum order parameter $\phi_{i}$ however counts beads of each type ($A$ or $B$) inside the bins and is not concerned with full molecules. Therefore those points with non-zero $\phi_{i}$ and null $P_{2}$ are drawn as empty circles within the partial free energy profiles.

On the other hand, when the system converges to its equilibrium starting composition and there is not splitting, a global approach may be used to identify the structure of different LC phases using a suitably defined order parameter. The isotropic and nematic phases can be characterised by defining the usual tensor $Q$ 
\begin{eqnarray}
Q \equiv \frac{1}{2N}\sum^{N}_{i=1}(3\hat{\textbf{u}_{i}}\otimes\hat{\textbf{u}_{i}} - \textbf{1})
\end{eqnarray}
where $\otimes$ is the dyadic product and $\textbf{1}$ is a unit tensor and the summation is taken over all the rod-like mesogens. The unit vector $\hat{\textbf{u}_{i}}$ points along the backbone of the rod like mesogens and is defined as the vector spanning the first and last beads $x^{(i)}_{1}-x^{(i)}_{N_A}$ for an arbitrary molecule $i$. The global nematic order parameter $S$ corresponds to the largest eigenvalue of the tensor $Q$, such that $S\approx0$ in the I phase and $S\approx1$ in the nematic phase ($N_{1}$) where molecules are aligned parallel to the nematic director $\hat{\textbf{n}}$.  The eigenvector associated with the largest eigenvalue is the global nematic order parameter $S$ and therefore contains information about the orientational ordering of molecules.

In order to probe the long ranged positional ordering in the smectic-A phase ($S_{1}$) and the distributions of the centre of mass of the rod-like mesogens along $\hat{\textbf{n}}$, the smectic order parameter must be introduced. It is given by the leading coefficient of the Fourier transform of the local density $\rho(\textbf{r}_{i}\cdot\hat{\textbf{n}})$. 
\begin{eqnarray}
\Lambda \equiv \frac{1}{N}\Bigg\langle\Bigg|\sum^{N}_{i=1}\exp\Bigg[\frac{2\pi i (\textbf{r}_{i}\cdot\hat{\textbf{n}})}{d}\Bigg]\Bigg|\Bigg\rangle
\end{eqnarray}
where $d$ represents the spacing between layers of rod-like molecules in a perfect Sm-A phase. This is predetermined to be $8.2\sigma$ from the density waves discussed in Figure \ref{fig:figure3} (d) for the $\phi_{0}=0.9$ composition. In a pure system of rod-like molecules, i.e. $\phi_{0}=1$, one might reasonably expect a perfect Sm-A phase to form, such that $d\approx k$ and $\Lambda=1$ but in the systems considered here this is rarely the case due to thermal fluctuations and the long run-times required to achieve perfect ordering. It is clear that any non-zero value indicates some degree of smectic ordering as evidenced by the density waves and snapshots, $\Lambda=0.1$ is therefore taken as a reasonable cutoff. By combining our new method, with the local and global approaches discussed here, it becomes possible to both map out the phase boundaries and characterise them. This is demonstrated in Section \ref{S3:Results_MFTvsMD} where the phase diagram is built from our CGMD simulations.

\section{Results \& Discussion}
\subsection{\label{S3:Results}Molecular Dynamics Simulations}

\begin{figure}[h]
\includegraphics[width=\columnwidth]{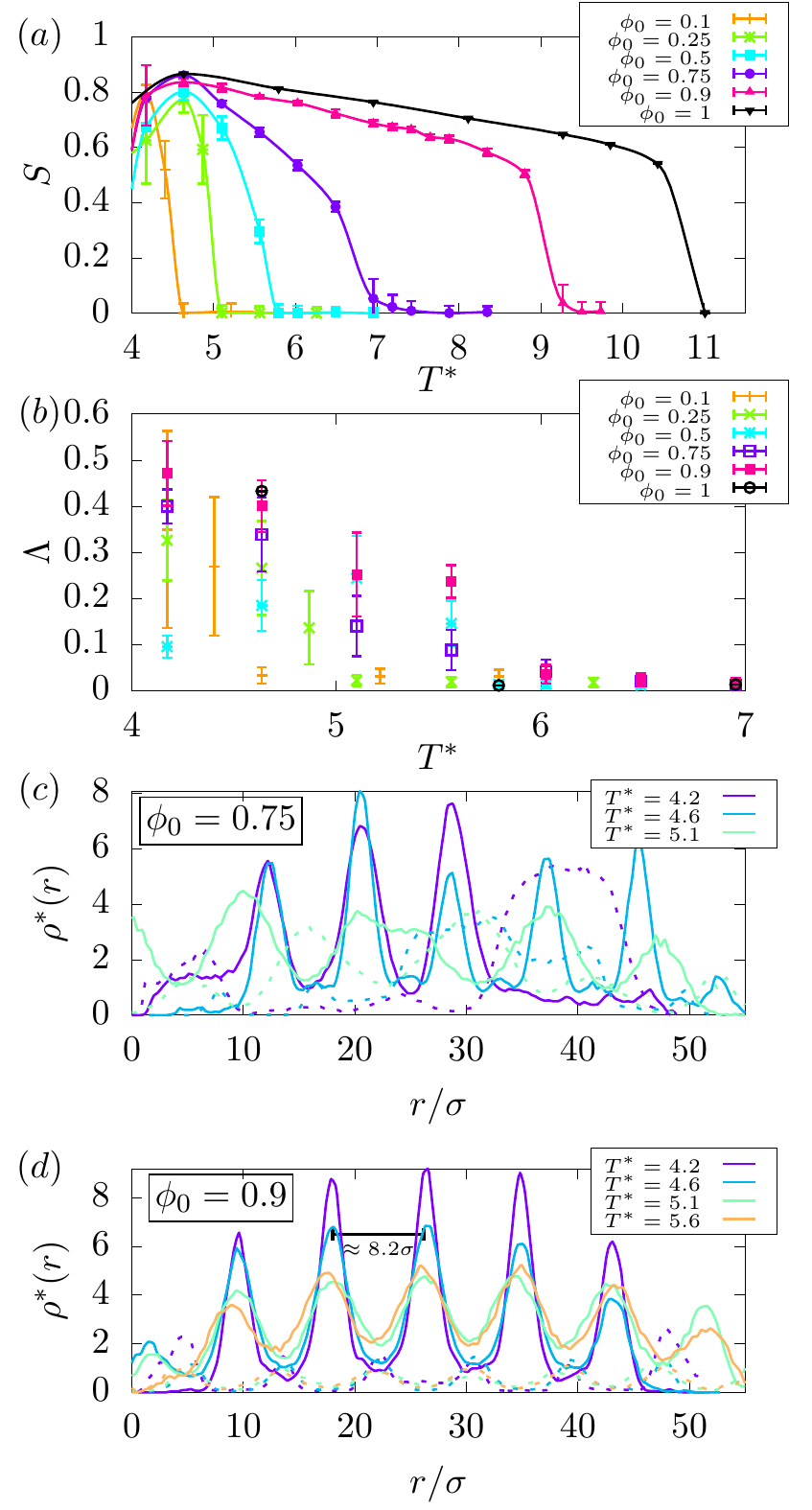}
\caption{\label{fig:figure3} (a) and (b) global nematic $P_{2}$ and smectic order $\Lambda$ parameters for each composition from MD simulation vs temperature. Both order parameters are calculated by averaging over the last 20ns of 5 independent quenches at each temperature. (c) and (d) density waves in the smectic phase for $\phi_{0}=0.75$ and $\phi_{0}=0.9$ compositions obtained by binning along the nematic director and averaging waves over the last 5ns of a single quench. The solid and dashed lines indicate the densities of the rod-like mesogens and flexible polymers respectively.}
\end{figure}  

Before analysing splitting and phase separation we first discuss the global observations from our CGMD simulations. Five separate initial LC volume fractions were considered in this study: $\phi_{0}=0.1,0.25,0.5,0.75$ and $0.9$ where the freely flexibly polymer and semi-flexible rod-like mesogen species had fixed lengths $N_{A}=4$ and $N_{B}=8$ bead units respectively. The global nematic and smectic order parameters, for all the compositions considered in this work, are shown in Figures \ref{fig:figure3} (a) and (b) respectively and have been averaged using the last 20ns of 5 independent quenches. It is apparent, from Figure \ref{fig:figure3} (a), that the nematic-liquid transition temperature $T_{NL}$, increases monotonically with an increasing LC volume fraction, towards the bulk nematic-isotropic transition temperature $T_{NI}$. This corresponds to the drop in the $\phi_{0}=1$ composition (black line), for a pure system of rod-like mesogens at around $T^{*}=10.5$, which was estimated by melting the pure system of rods. The $\phi_{0}=1$ composition exhibits LC phases on its own, at the lowest temperatures $T^{*}<5.5$ the $S_{1}$ phase appears first in which layers of aligned rods stack on top of one-another with well defined spacing, where $\Lambda \approx 1$. This gradually decreases until $T^{*}\approx5.5$ at which point long-range positional order is lost ($\Lambda\approx0$) and the $N_{1}$ phase appears where the rods retain rotational order, $S\approx0.8$. As the temperature is raised the nematic order continues to decrease towards a value of $S\approx0.75$ until $T_{NI}=10.5$ at which point all rotational order is lost and the system is completely isotropic. This behaviour has been observed in a number of similar studies of rod-like mesogens \cite{milchev2019smectic} and is not unexpected. Aside from the pure system, the remaining compositions with a flexible polymeric component, were studied upon quenching the system from the isotropic phase at $T^{*}=10.5$ ($T_{NI}$) to ensure that no rotational ordering of the mesogens remains in any of the compositions studied for $\phi_{0}\leq0.9$. 

For $T^{*}\geq6.0$ the compositions with a large volume fraction of mesogens $0.75\leq\phi_{0}\leq1$ are clearly nematic ($N_{1}$) with $S\approx0.8$ and $\Lambda\approx0$. Compositions with $\phi_{0}<0.75$ are completely isotropic ($L_{1}$) with $S\approx0$. As temperature is further decreased to $T^{*}=5.6$ compositions with $\phi_{0}\geq0.5$ show a non-zero $S$ and $\Lambda$ indicating the presence of both rotationally ordered and positionally ordered regions. We speculate that $T^{*}=5.6$ is close to the point where $S_{1}$, $N_{1}$ and $L_{1}$ phases may coexist \cite{mukherjee2020wetting}. Even though the smectic ordering retains only a small non-zero value ($\Lambda\approx 0.25$) for the $\phi_{0}=0.9$ composition, it is clear from Figure \ref{fig:figure3} (d) that there is preferential ordering of the rod-like mesogens into bands, with the flexible polymers filling the interstitial regions indicated by the solid and dashed lines respectively. Those compositions with compositions lying between  $0.5\leq\phi_{0}\leq0.75$ also show a non-zero $\Lambda$ indicating small $S_{1}$ domains may exist. All other compositions $\phi\leq0.25$ are completely isotropic at this temperature. 

In the region where $4.2\leq T^{*} \leq 4.6$ the smectic ordering $\Lambda$ for compositions $0.5\leq\phi_{0}\leq1$ gradually increases accompanied by an increased $S$ indicating a more ordered and micro-phase separated $S_{1}$ phase appears. This is no more apparent than in Fig \ref{fig:figure3} (c) and (d) where the mesogen-rich regions contain no flexible polymers in comparison to higher temperatures $T^{*}>4.6$ as well as a reduction in the number of mesogens in the polymer-rich regions. Importantly for the $\phi_{0}=0.9$ composition, the system fully adopts the $S_{1}$ phase as seen in Figure \ref{fig:figure3} (d) whereas the $\phi_{0}=0.75$ composition always contains regions with what appears to be some splitting with a low density liquid phase. This is shown most clearly at $T^{*}=4.2$ in Figure \ref{fig:figure3} (c) where the system is split between the $S_{1}$ phase and with 3 clearly defined peaks in one half of the simulation cell, with the other side containing a small number of rod-like mesogens dispersed in the flexible polymers. This should feature prominently in the MD phase diagram; the absence of the $N_{1}$ phase would also suggest that $T^{*}=4.6$ is below the triple point where only $S_{1}$ and $L_{1}$ phases may coexist. Similar self-organisation is also evident in experimental systems of 
binary mixtures of long and short PDMS molecules, where they phase-segregate into alternate layers of long and short smectic phase owing to entropic stabilisation \cite{okoshi2010alternating,kato2019smectic}. Here, we observe similar micro phase-separated phases owing to combined effects of entropy and enthalpy.

\subsection{\label{S3:Results_MFTvsMD}Phase Diagrams}
\begin{figure}[htb]
\includegraphics[width=\columnwidth]{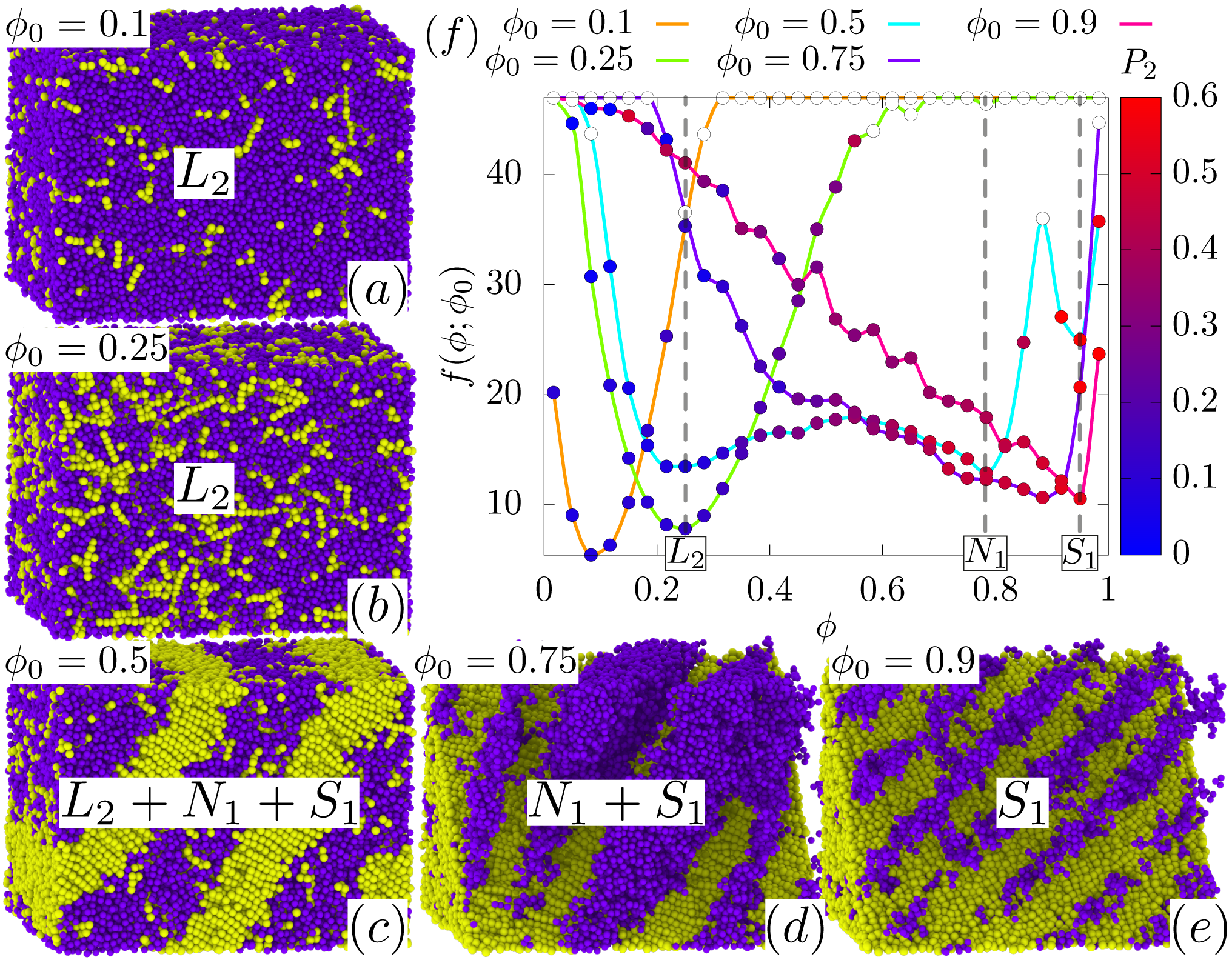}
\caption{\label{fig:figure4} Constructing partial free energy landscapes from MD simulations at $T^{*}=5.1$. (a-e) Snapshots taken from MD simulations as the LC component is increased, the flexible polymers and semi-flexible rod-like mesogens are coloured purple and yellow respectively to enhance their orientational alignment and in panels (d) and (e) half the rods have been removed to reveal the banding of purple polymers in the smectic phase, produced using OVITO \cite{stukowski2009visualization}. (e) Free energy profiles as inverted from the probability distributions in Figure \ref{fig:figure2} (c), the approximate locations of the phase boundaries are indicated by dashed (grey) lines. The points along the histogram have been coloured continuously, according the local nematic order parameter $P_{2}$ of the rods, as indicated by the colourbar on the RHS, from $P_{2}=0$ (blue) to $P_{2}=0.6$ (red). In this way it is possible to distinguish between isotropic and anisotropic minima in the free energy landscape.}
\end{figure}
\begin{figure*}[htb]
\includegraphics[width=\textwidth]{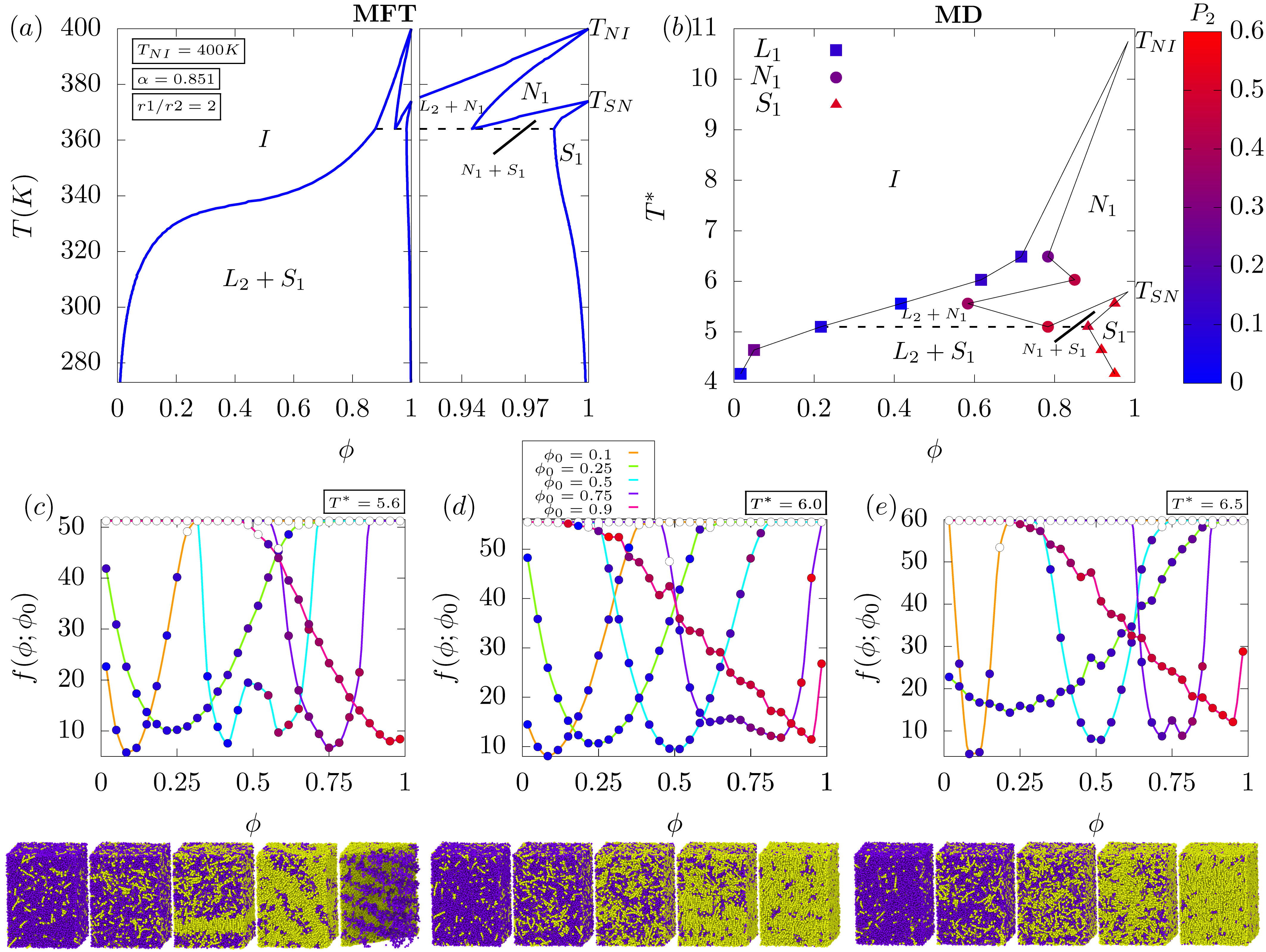}
\caption{\label{fig:figure5} Phase diagram of a binary polymer-smectic-liquid crystal mixture. (a) Phase diagram as calculated from the mean-field theory with parameters as indicated in the figure in the long rod regime, see Appendix \ref{Appendix:MFT} for a detailed methodology of the numerical procedure. (b) Phase diagram as extracted from CGMD simulations, point types correspond to different phases: $\blacksquare$-Liquid, $\CIRCLE$-Nematic and $\blacktriangle$-Smectic. Each point is coloured according to the local nematic order $P_{2}(\phi)$, with the corresponding value in the colourbar (rhs). (c-d) Effective free energy profiles $f(\phi;\phi_{0})$ extracted from CGMD simulations at all compositions considered for $T^{*}=5.6$, $6.0$ and $6.5$ where the points are coloured according to their local $P_{2}$ values according to the colour scale in panel (b). Shown alongside are the corresponding snapshots for each of the compositions considered at each temperature, the flexible polymers and semi-flexible rod-like mesogens are coloured purple and yellow respectively.}
\end{figure*}

  
For reference a brief recap of the theoretical model for predicting phase diagrams of mixtures of polymers and smectic liquid crystals is presented and the key parameters which govern the phase behaviour are described.  More details about the model and its development can be found in refs \cite{mcmillan1971simple,kyu1996phase,chiu1998phase} and Appendix \ref{Appendix:MFT}. The free energy of a mixture of a polymer and a smectic liquid crystal $f=f_{iso}+f_{aniso}$ comprises of two parts, an isotropic part describing the thermodynamics of isotropic liquids $f_{iso}$ and an anisotropic part which accounts for the ordering of the liquid crystals $f_{aniso}$.  Flory-Huggins theory, \cite{flory1953principles} is used to describe the former for a liquid crystal - polymer mixture such that
\begin{eqnarray}
\scalebox{0.95}[1]{$f_{iso}(\phi,T) = \frac{\phi}{r_{1}}\ln\phi  + \frac{1-\phi}{r_{2}}\ln(1-\phi) + \chi(T) \phi(1-\phi)$}
\label{e:fh}
\end{eqnarray}
where $r_{1}$ is the length of the rod-like mesogens, $r_{2}$ is the length of the polymer and $\phi$ is the volume fraction of the LC component. The F-H interaction parameter $\chi(T)$ is a quantity accounting for the enthalpic interactions and it is generally described by an inverse temperature relationship of the form $\chi=A+\frac{B}{T}$, where $A$ and $B$ are material specific parameters. The anisotropic portion, which couples the LC composition into the free energy is given by 
\begin{eqnarray}
f_{aniso}(\phi,T,m_{n},m_{s}) = -\Sigma(m_{n},m_{s}) \phi \\*
\nonumber
- \frac{1}{2}\nu(T)(s(m_{n})^{2}+\alpha \kappa(m_{s})^{2})\phi^{2} 
\label{e:msm}
\end{eqnarray}
where $\Sigma$ represents the decrease in entropy as the rod-like polymers align (Equation \ref{SI_MFT_ENTROPY}), $s$ is the nematic order parameter (Equation \ref{SI_MFT_NEMATIC}) and $\kappa$ is the smectic order parameter (Equation \ref{SI_MFT_SMECTIC}). All of these quantities are functions of the dimensionless nematic and smectic mean-field terms $m_{n}$ and $m_{s}$. The nematic coupling term $\nu(T)$ is a temperature dependent term which depends on the nematic-isotropic transition temperature $T_{NI}$, such that $\nu(T)=4.541 T/T_{NI}$, note the pre-factor is a universal quantity \cite{kyu1996phase,chiu1998phase}. The smectic interaction coupling $\alpha$, is a dimensionless quantity as defined in Equation \ref{SI_MFT_ALPHA} and is kept fixed. By minimising the free energy functional with respect to the order parameters ($\frac{\partial f_{aniso}}{\partial s}=0$ and $\frac{\partial f_{aniso}}{\partial \kappa}=0$), the resulting expressions (Equations \ref{SI_SELFCON_1} and \ref{SI_SELFCON_2}) can be evaluated numerically using the procedure outlined in Appendix \ref{Appendix:MFT}. The renormalised free energy landscape (obtained after re-substituting the minimised values of the nematic and the smectic order parameters back into the full free energy expression) is then determined at a given temperature (see Figure \ref{fig:figure9}) and the phase diagram can be mapped out as illustrated in Figure \ref{fig:figure5} (a). This is discussed in conjunction with the phase diagram extracted from our CGMD simulations.  

Figure \ref{fig:figure5} (b) shows the binary phase diagram for the semi-flexible rod-like mesogens with stiffness constant $k_{bend}=50$ and $N_{A}=8$ and fully-flexible polymers with $N_{B}=4$ as extracted from our CGMD simulations. This has been reconstructed using the procedure outlined in Section \ref{S2:Method} such that the local minimums of $f(\phi;\phi_{0})$, that result from the splitting of the effective free energies or order parameter distributions, are taken to define the phase boundaries. For $T^{*}<5.1$ the system shows a pure liquid phase $L_{2}$ and pure smectic-A phase $S_{1}$, at very low and very high mesogen concentrations respectively. This is evidenced by the value of the local $P_{2}$ order parameter in both regions as highlighted by the colouring of the points in Figure \ref{fig:figure4} (f) and a non-zero value of the global smectic ordering $\Lambda$ in Figure \ref{fig:figure3} (b). Between these two regions lies a large $L_{2}+S_{1}$ coexistence region spanning the intermediate concentrations.

At $T^{*}=5.1$, $L_{2}$ moves inwards to higher concentrations ($\phi\approx0.25$) and $S_{1}$ moves similarly to lower concentrations ($\phi\approx0.9$) and another highly ordered nematic phase appears $N_{1}$, this demarcates the $L_{2}+N_{1}$ and $N_{1}+S_{1}$ coexistence regions. The effective free energy profiles at $T^{*}=5.1$ are shown in Figure \ref{fig:figure4} (f) where the $\phi_{0}=0.5$ simulation (cyan line) shows the 3 distinct regions corresponding to the $L_{2}$, $N_{1}$ and $S_{1}$ phase boundaries. This is further confirmed by the local $P_{2}$ ordering in all 3 regions with $L_{2}$ phase corresponding to $P_{2}\approx 0$ and the $N_{1}$ and $S_{1}$ phases reaching $P_{2}\approx 0.5$ showing that they are highly ordered. In order to distinguish these two ordered minima we consult Figure \ref{fig:figure3} (a) and (b) and note that at high concentrations $\phi_{0}>0.5$ which show this splitting have both $\Lambda \neq 0$ and $S \neq 0$ indicating the presence of nematic and smectic phases. However this is not enough to identify which minimum corresponds to the smectic phase since a measure of the local smectic ordering or indeed its distribution is not possible to calculate, unlike the local nematic ordering ($P_{2}$), see Figure \ref{fig:figure4} (f). We therefore examine the simulation snapshots in Figure \ref{fig:figure2} taken at $T^{*}=5.1$ which shows the $\phi_{0}=0.5$ concentration (Figure \ref{fig:figure2} (c)) is predominantly split between $L_{1}$ at low concentrations and $N_{1}$ at high concentrations, whereas the snapshots for the $\phi_{0}>0.5$ concentrations (Figure \ref{fig:figure2} (d-e)) are predominantly $S_{1}$. 

As temperature is further increased to $T^{*}=5.6$ both $L_{2}$ and $N_{1}$ appear to move inward
with only the highest composition $\phi_{0}=0.9$ having any smectic ordering with $\Lambda>0$, see Figure \ref{fig:figure3}(b). At $T^{*}=6.0$ all positional ordering is lost and the $S_{1}$ phase is replaced by the $N_{1}$ phase, this is the first spout of the double ``teapot'' topology indicated by $T_{SN}$ in figure \ref{fig:figure5} (b). The $L_{2}$ region moves to even higher $\phi$ values leaving a narrow $L_{1}+N_{1}$ coexistence region as evidenced by the splitting of the $\phi_{0}>0.75$ simulation in Fig \ref{fig:figure5}(d). This region narrows further at $T^{*}=6.5$ in Figure \ref{fig:figure5}(e) after which the minimums appear to merge with no apparent splitting at higher temperatures. However this does not mean that the coexistence region is lost but instead that it has narrowed sufficiently such that the compositions at which simulations have been performed do not fall inside this narrow coexistence region. We speculate that this region could be isolated by considering compositions in the region $0.75\leq\phi_{0}\leq0.9$, it is sufficient however to simply connect these minimums to the nematic-isotropic transition temperature $T_{NI}$ determined by melting the pure system ($\phi_{0}=1$). This give the second spout of the double ``teapot'' topology indicated by $T_{NI}$ in figure \ref{fig:figure5} (b).

From the MFT we find a similar picture to that extracted from the CGMD simulations, this is shown in Figure \ref{fig:figure5} (a) where two crucial changes have been made to the parameter values originally presented in \cite{kyu1996phase}, see Figure \ref{fig:figure1} or \ref{fig:figure10} (a) for the original phase diagram (for a detailed discussion on how the shape of the phase-boundary is affected by the parameter values refer to Section 4 of Appendix C). Firstly the polymer:mesogen length ratio $r_{2}/r_{1}=2$ has been inverted such that the mesogens are twice as long as the polymers to match our CGMD simulations. This has the effect of pushing the original $L_{1}+L_{2}$ coexistence region to lower concentrations as well as suppressing the temperature at which these two phases merge $T_{c}$, see Figure \ref{fig:figure10} (a) and (c). Secondly the nematic-isotropic transition temperature $T_{NI}$ has been raised from 333K to 400K, motivated by the high stiffness of our mesogens in our CG simulation model, which raises the temperature of the melt. This has the effect of pushing the double chimney-like topology upwards in the phase diagram to higher temperatures. As a direct consequence the $L_{1}+L_{2}$ coexistence region then becomes buried inside the phase diagram and is replaced by the $L_{1}+S_{1}$ region which also moves outward such that the pure $L_{2}$ and $S_{1}$ phases are forced to extremely low and high concentrations respectively due to an increased $T_{NI}$. Thus we observe $L_{2}+S_{1}$, $N_{1}+L_{2}$ and $N_{1}+S_{1}$ regions as well as pure $N_{1}$ and $S_{1}$ regions but no $L_{1}+L_{2}$ region. This bears a similar resemblance to the phase diagram obtained from our model CGMD simulations, possessing identical qualitative features. 

\section{\label{S4:Conc}Conclusions}

A new methodology has been developed to probe the topology of free energy landscapes, from CGMD simulations of binary mixtures, by manipulating continuum order parameter distributions. Using our method we have shown how the approximate locations of the phase boundaries (spinodals) can be extracted and then characterized by analysing global nematic and smectic order parameters, local nematic order parameter distribution and simulation snapshots. The resulting phase diagram was then compared with Maier-Saupe type mean-field theory using comparable parameters to our MD simulations. Both diagrams possess an identical double chimney-like topology, even with modest computational resources, demonstrating the power of this method.

The accuracy of our method has a strong dependence on the shape of the phase diagram, specifically the width of the region in $\phi$ space. If the region is sufficiently wide, it is more likely that one of the initial starting compositions $\phi_{0}$, from our MD simulations, will fall inside the region and splitting will be observed. For our regime, long rods and short polymers, the $S_{1}$, $N_{1}$, $L_{1}+N_{1}$ and $N_{1}+S_{1}$ regions appear at very high volume fractions of the LC component and are narrow. In addition as the temperature is raised these regions further narrow considerably which hinders the method at higher temperature. In the reverse scenario however, with short rods and long polymers, an additional $L_{1}+L_{2}$ region is present. This region would be more accurately probed by our method since the different phases are more well separated in $\phi$ space. 

Additionally, a quantitative estimate of $\chi(T)$ from the temperature dependence of the location of the peaks of the order parameter distribution is a relatively simple exercise for an isotropic polymer mixture (see the isotropic $L_{1}+L_{2}$ coexistence region in the phase diagram of short rods and long polymers) described by a Flory-Huggins free energy. This estimate, however, gets more involved as one includes anisotropic phases in the description, especially for systems with long mesogens, as in the systems consided here. The anisotropic phases start appearing at even lower volume fractions and interfere with the Ising like critical point. Here one observes the effects of the interference of the discrete Ising symmetry associated with the $\phi$ order parameter and the continuous symmetry of the nematic and the smectic order parameters and this makes the quantitative estimate of $\chi(T)$ more difficult. This is an aspect associated with the exact matching of the phase diagram resulting from MD simulations to that from the MFT, which would be resolved in future.

A computational method like this is general enough to be applied to the sub-cellular environment where
semi-flexible bio-polymers undergo liquid-liquid phase separation and under specific physico-chemical conditions they can self-assemble into non-random filamentous structures with anisotropic interactions promoting nematic ordering. It is also known that mechanical strain may induce alignment of semi-flexible polymers. Thus a method like this becomes an important tool for estimating parameters ($\chi(T)$, $\nu(T)$, $T_{NI}$ etc.) for constructing phase diagrams which thus enables a realistic meso-scale description of specific bio-polymers which already accounts for the specific chemical details. This specific meso-scale model can also be used for non-equilibrium kinetic simulations where one can probe the important role of various metastable intermediates in these complex systems.

\begin{acknowledgments}
W.F., B.M. and B.C. acknowledge funding support from EPSRC via grant EP/P07864/1, and P\& G, Akzo-Nobel, and Mondelez Intl. Inc. This work used the Sheffield Advanced Research Computer (ShARC) and BESSEMER HPC Cluster.
\end{acknowledgments}

%Appendices
\appendix
\section{\label{Appendix:Methods_Rationalise}A Rationalisation For Reconstructing Free Energy Landscapes}

In order to rationalise our method of guessing the nature of the free energy landscape by monitoring order parameter distributions at various compositions and finally combining them, we have performed some ``model" computations.
We simulate a conserved-order parameter dynamics (model B) on a (100 $\times$ 100) square lattice and the 
dynamic concentration profiles for the phase-separation order parameter, $\phi({\bf r}, t)$, satisfies 
\begin{eqnarray}
\frac{\partial \phi({\bf r},t)}{\partial t} = \nabla \cdot \left[ M \nabla \frac{\delta F[\phi({\bf r},t)]}{\delta \phi({\bf r},t)} + \theta({\bf r}, t) \right], 
\label{e:diff_eqn}
\end{eqnarray}
where $M$ is the mobility, assumed to be composition independent and the local chemical potential $\mu(\phi({\bf r}, t)) = \frac{\delta F[\phi({\bf r},t)]}{\delta \phi({\bf r},t)}$. An additive vectorial  conserved noise $\theta({\bf r}, t)$ in Eq.\ref{e:diff_eqn} modelling solvent effects, satisfying $\langle \theta_{i}({\bf r}, t) \rangle$ = 0, and $\langle \theta_{i}({\bf r}, t) \theta_{j}({\bf r^{\prime}}, t^{\prime}) \rangle = 2 M k_{B} T \delta_{ij} \delta ({\bf r} - {\bf r^{\prime}}) \delta (t - t^{\prime})$ ensures thermodynamic equilibrium at long times. The free energy functional for an in-compressible binary fluid mixture, in two space dimensions, is given by 
\begin{eqnarray}
F[\phi({\bf r})]/k_{B}T = \int_{0}^{d} \int_{0}^d [f(\phi) &+& k (\nabla \phi)^2 ] dx dz, 
\label{e:free_en_func}
\end{eqnarray}
where $F$ is the free energy, and $z$ and $x$ are the spatial coordinates. The first term in Eq.~\ref{e:free_en_func} is the bulk free energy and the second term accounts for energy costs associated with the spatial gradients of the composition field with a stiffness coefficient 
$k$.

\begin{figure}[h]
\begin{center}
\includegraphics[]{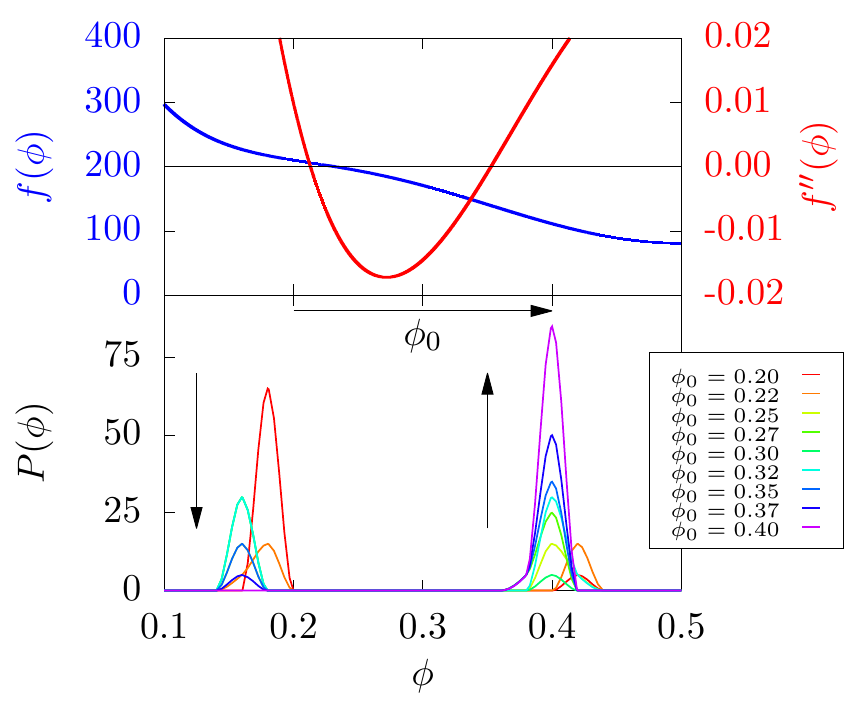} 
\caption{The model three minimum free energy (blue) and the regions of 
positive and negative curvatures (red) (upper panel). The order parameter histograms evaluated at long times for various values of $\phi_{0}$ (lower panel).}
\label{fig:figure6}
\end{center}
\end{figure}

For the free energy f($\phi$) we choose a model free energy (upper panel of Fig. \ref{fig:figure6}) with multiple minima and regions of both positive and negative curvatures (lower panel of Fig. \ref{fig:figure6}) present in the free energy landscape. We simulate the temporal evolution of systems initiated from various $\phi_{0}$ (plus a delta correlated noise term with an amplitude of 0.05) via Eq. \ref{e:diff_eqn} and monitor the order parameter distribution at long times. In this computation we have used $\Delta x$ = $\Delta z$ = 0.5 and $\Delta t$ = 10$^{-6}$ for the spatial and temporal discretisation and simulation for each $\phi_{0}$ has been performed for 10$^{8}$ time-steps and the order parameter distributions have been computed from the final configuration. We observe that when the simulation is initiated from a $\phi_{0}$ for which f$^{\prime \prime}$($\phi$) is negative, the long time order parameter distribution (see Fig. \ref{fig:figure6}) shows a split peak at two values of $\phi$ which are the extremities of a common-tangent bracketing the initial unstable $\phi_{0}$. On the other hand, when initiated from a stable $\phi_{0}$ the long time order parameter distribution is a gaussian centred at $\phi_{0}$. Thus the nature of order parameter distribution for various $\phi_{0}$ allows us to map out the essential features of the free energy landscape.

\section{\label{Appendix:CGMD_Model}Coarse-Grained MD Model Details}
A brief overview of our model \cite{mukherjee2020gelation}, used for both the polymer (A) and liquid crystal species (B), is given in this section for reference, including how it is extended to control the rigidity of the model mesogens. The bonded interactions between coarse-grained beads for both species are described using the FENE potential \cite{kremer1990dynamics},
\begin{eqnarray}
U_{bond}=-\frac{1}{2}k_{bond}r_{0}^{2}\log\Bigg[1-\Big(\frac{r}{r_{0}}\Big)^{2}\Bigg]
\end{eqnarray}
where $U_{bond}$ is the change in potential energy associated with bond stretching, $k_{bond}$ is the spring constant and $r_{0}$ is the bond distance or range of the bond potential, see Table \ref{T1:Bonds} for a list of parameter values.

\begin{table}[t]
\caption{\label{T1:Bonds}Bond parameters for MD, note values chosen for $\sigma=0.339$ nm, while $\epsilon=0.359$ kJ/mol and the value of masses of all the beads have been chosen as $m=12.01$ amu.}
\begin{ruledtabular}
\begin{tabular}{cccc}
Type & $k_{bond}$ ( $\epsilon/\sigma^{2}$) & $r_{0}$ ($\sigma$) & $k_{bend}$\\
\colrule
A & 40 & 1.5 & 50\\
B & 40 & 1.5 & 0\\
\end{tabular}
\end{ruledtabular}
\end{table} 

Additional rigidity was also included via a bending potential where each set of three consecutive beads along the mesogens backbone interact via a harmonic potential,
\begin{eqnarray}
U_{bend}=k_{bend}(1-\cos \theta)
\end{eqnarray}
where $U_{bend}$ is the potential energy change associated with the change in bond angle from its equilibrium position and $k_{bend}$ is the angle spring constant, related to the persistence length $\frac{l_{p}}{l}=\frac{k_{bend}}{k_{B}T}$.
Non-bonded interactions between like and unlike beads interact via pairwise 12-6 Lennard-Jones potentials of the form,
\begin{eqnarray}
U_{LJ_{12-6}}=4\epsilon_{\alpha\beta}\Bigg[ \Big(\frac{\sigma}{r}\Big)^{12} - \Big(\frac{\sigma}{r}\Big)^{6} \Bigg]
\end{eqnarray}
where $r$ is the distance between pairs of beads and the indices $\alpha$ and $\beta$ denote the binary species. In order to ensure phase-separation the species dependent term $\epsilon_{\alpha\beta}$, is chosen such that $\epsilon_{AA}=\epsilon_{BB}=2\epsilon_{AB}$. We note that variable persistence length alone has been demonstrated recently to be sufficient in itself to facilitate entropic un-mixing in similar systems \cite{milchev2020entropic}.

Throughout lengths, times and temperatures are expressed as dimensionless quantities such that $l^{*}=l/\sigma$, $\tau^{*}=\tau/\sqrt{m\sigma^{2}/\epsilon}$ and $T^{*}=k_BT/\epsilon$ respectively.  Each composition was prepared such that the dimensionless density $p^{*}=N\sigma^{3}/(L_{x}L_{y}L_{z})\approx 1$ to ensure a liquid system far from solid-liquid and liquid-gas transitions at the simulated temperatures. Initial configurations were prepared by performing simulations for $\tau=2\times10^{5}$ timesteps at $T^{*}=10$ for each composition and extracting 5 independent starting configurations. These configurations were then instantaneously quenched to a series of temperatures between $T^{*}=10$ and $T^{*}=3$ at $\Delta T^{*}=0.4$ intervals. Variable simulation times were used since simulations at lower temperatures, although quick to phase-separate, take longer to order than those at higher temperatures which equilibrate fast. Simulations in temperature intervals $4.2<T^{*}\leq5.5$, $5.5<T^{*}\leq7.0$ and $7.0<T^{*}<10.5$ were run for 160ns, 80ns and 40ns respectively. All simulations were performed in a constant volume ensemble and the temperature maintained by a Nose-Hoover thermostat.

%%%BROUGHT IN ORIGINAL SI INTO APPENDICES 

\section{\label{S1:Methods}Methods}
\subsection{\label{S1:Intro_Histograms}Numerical Methodology for Extracting Histograms}

\begin{figure}[htb]
\begin{center}
\includegraphics[width=\columnwidth]{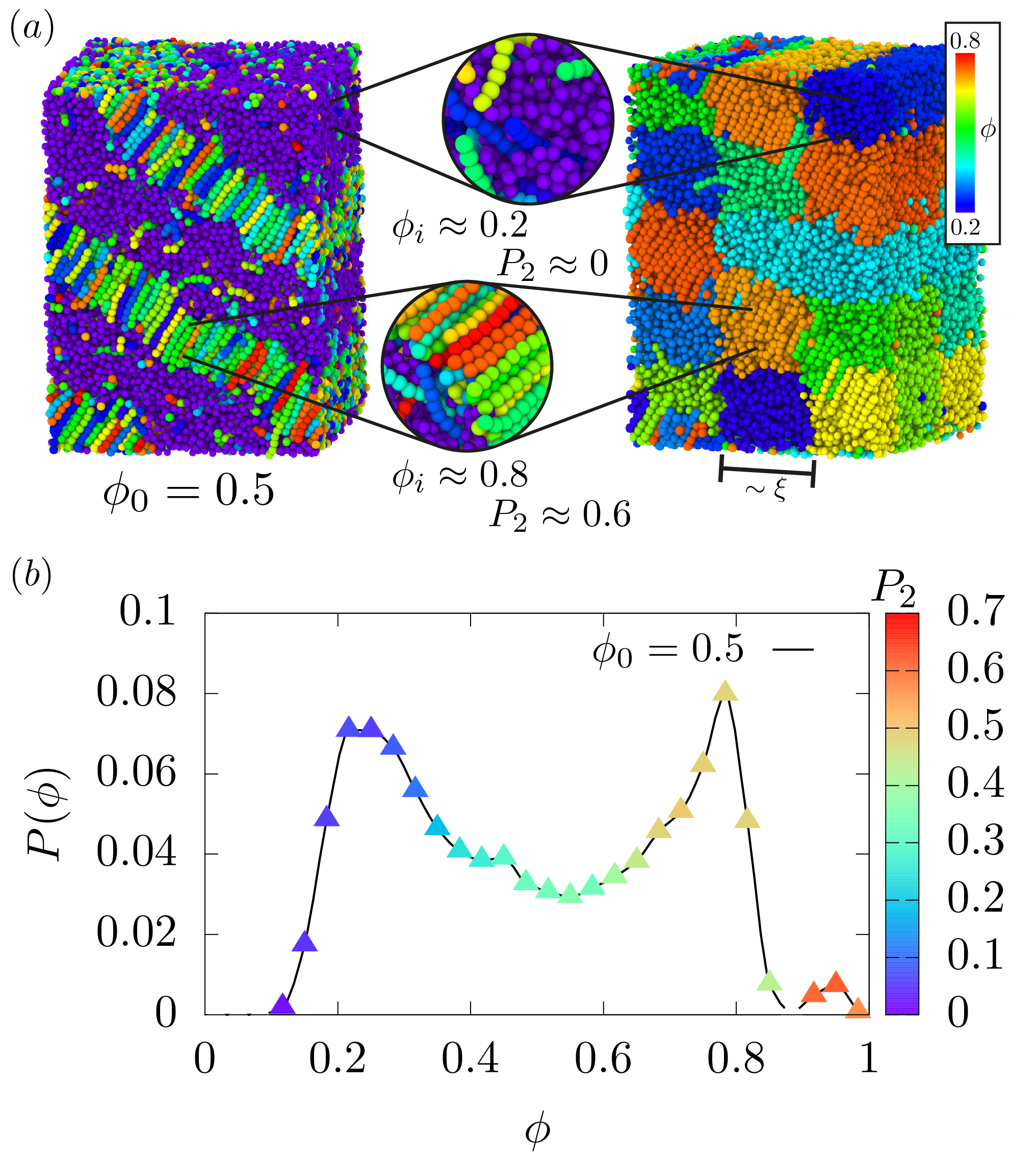} 
\caption{(a) Snapshot of the $\phi_{0}={0.5}$ configuration at $T^{*}=5.1$ showing coexisting liquid and nematic phases (left) and the simulation cell as binned where the molecules in the cells are coloured according to their local composition (right). The bin size which is comparable to the correlation length has been drawn in. (b) Probability distribution for the $\phi_{0}=0.5$ composition where each point is coloured according to local order parameter corresponding to the colour bar (right). In the compartment snapshots the rod-like mesogens are coloured randomly and the flexible polymers are purple. Note at this temperature the low-$\phi$ maximum in $P(\phi)$ is identified as liquid due to its low $P_{2}$ values whereas the high-$\phi$ maxima are identified as liquid crystalline (nematic) due to their high $P_{2}$ values.}
\label{fig:figure7}
\end{center}
\end{figure}

Using the correlation length $\xi$, as determined from the zero-crossing of the $C(r_{ij})$ profile described in the main manuscript, the simulation cell is re-binned into cubes with dimensions $\approx (\xi)^3$. This is depicted in Figure \ref{fig:figure7} (a) where the bin size, comparable with the correlation length, has been drawn in and the molecules inside each of the cells have been coloured according to their composition $\phi_{i}$ as defined in the main manuscript. Note the edges of the simulation cell have been cutoff to more cleanly show the boundaries between each of the binned compartments. This particular snapshot is taken close to the point at L, N and Sm-A phases may coexist at $T^{*}=5.1$ for the $\phi_{0}=0.5$ composition. The $P(\phi)$ distribution is then determined by counting the frequency of number of bins with compositions that fall into a certain $\phi$ interval and producing a histogram. The resulting histogram is shown in Figure \ref{fig:figure7} (b) and has been averaged over the last 20ns of all quenches. Some of the compartments with compositions corresponding to the peak values have been isolated and zoomed-in to illustrate both the arrangement of molecules inside the cells and the overall composition. 

In order to assess the local ordering of the rod-like mesogens inside each cell, their individual local $P_{2}$ values may also be averaged such that each compartment has both a corresponding density and $P_{2}$ value, see Appendix \ref{SI_P2} for the appropriate definition. The cutoff used for the local $P_{2}$ calculation corresponds to the HWHM in the $P_{2}(r)$ profile in the main manuscript. Then the $P_{2}$ values of those compartments used to produce each point along the $P(\phi)$ distribution can be isolated and averaged to give an estimate of the local alignment of molecules at different values along the $P(\phi)$ profile. This corresponds to the colour of the individual points in Figure \ref{fig:figure7} (b), it can be seen that the low-$\phi$ peak corresponds to a low density liquid phase, where the alignment of the rod-like molecules are almost completely random ($P_{2}<0.1$) and the remaining two peaks correspond to high density liquid crystalline phases ($P_{2}>0.5$). It is interesting to note that the smaller peak at around $\phi\sim0.95$ is more highly ordered ($P_{2}\approx 0.6$) than the more prominent peak at $\phi\sim0.8$, we speculate that this could be due to very few polymers entering the bands of the rod-like mesogens and that the separation between the nematic band and the surrounding flexible-polymers is more cleanly defined. This is characteristic of the layers in a Sm-A configuration hence we speculate this final peak is the smectic phase appearing, likely due to the close proximity to the triple point and thermal fluctuations.

\subsection{\label{S2:Intro_Order}Global Nematic Ordering $S$ and Local Nematic Ordering $P_{2}$}

\subsubsection{\label{SI_P2}Local Order Parameter}

\begin{figure}[htb]
\centering
\includegraphics[width=\columnwidth]{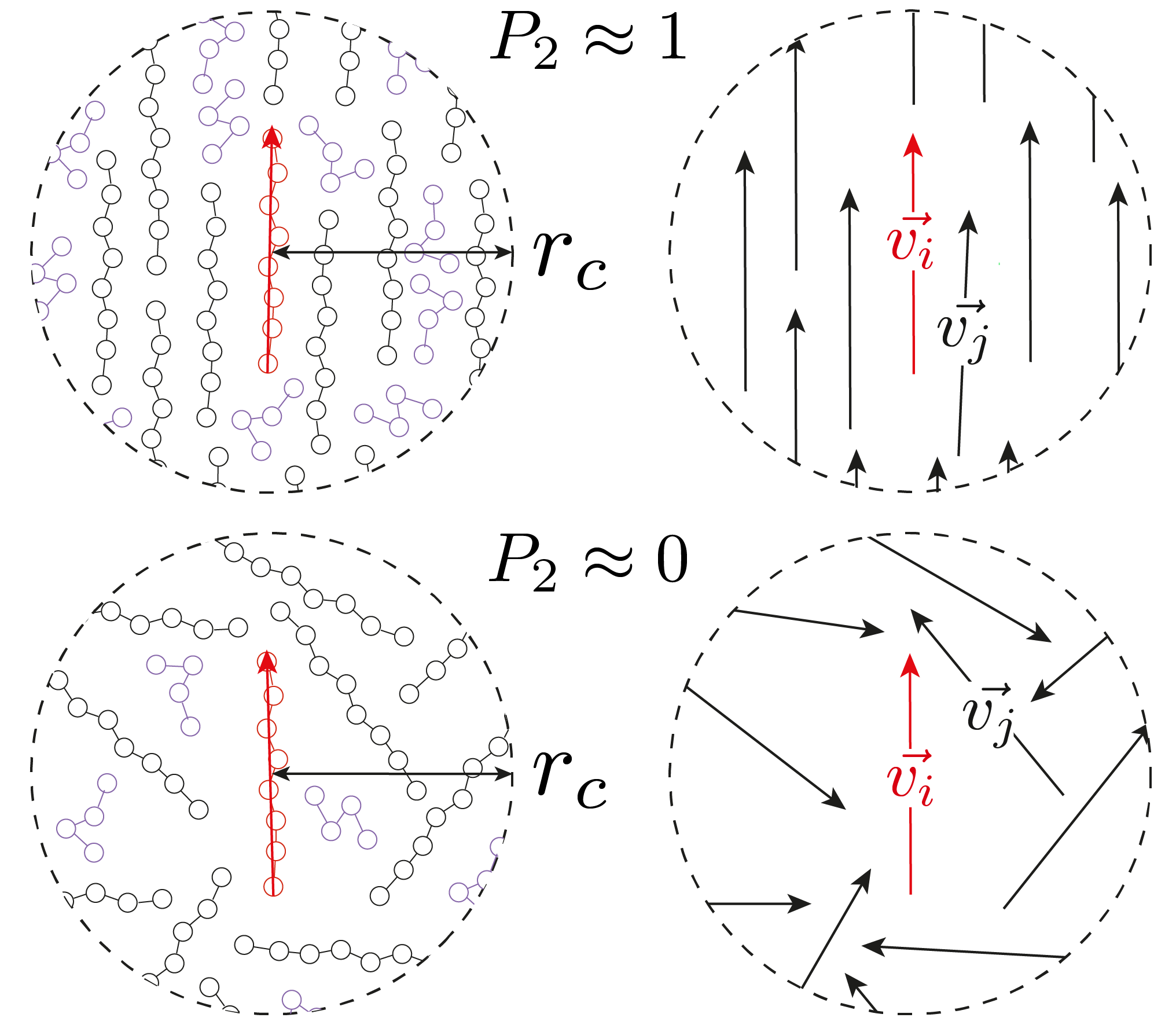}
\caption[]{Assigning the local ordering, $P_{2}$ order parameter, of a single semi-flexible rod $\vec{v_{i}}$ with the remaining black rods $\vec{v_{j}}$, inside the cutoff $r_{c}$ following the procedure outlined in \cite{fall2020statistical}. Rods are coloured black/red and the flexible polymers are drawn in purple and are neglected in the calculation. The first scenario (top) represents a highly ordered liquid crystalline configuration $P_{2}\approx 1$ where the semi-flexible rods are generally aligned with one another and the second where the rods are almost completely randomly oriented in a liquid such that $P_{2}\approx 0$. }
\label{fig:figure8}
\end{figure}

The local $P_{2}$ order parameter is a measure of the local alignment between rods within a given cutoff distance $r_{c}$. For an arbitrary rod $i$ and its neighbours ($j=1,...,N)$, its local $P_{2}$ ordering is given by
\begin{equation} 
P_{2}(i)=\frac{1}{N} \sum_{j=1}^{N}\frac{3\cos^{2}\theta_{ij} -1}{2}, \;\;\;\;\; r\leq r_{c}
\end{equation}
%\begin{equation}
%P_{2}_{i} =\frac{1}{N} \sum_{i=1}^{N}\frac{3\cos^{2}\theta -1}{3}
%\end{equation}
Where $\theta_{ij}$ is the angle between the backbone vector of rod $i$ with the $j$th rod inside the cutoff distance and $N$ is the number of neighbouring rods with a centre of mass that falls within the cutoff. The backbone vector of a given semi-flexible rod is approximated as the vector spanning the first and last beads of the rods such that $\vec{v_{i}} = x_{1}-x_{N_{A}}$ as indicated by the (red) arrow in Figure \ref{fig:figure8}. In Figure \ref{fig:figure8} the angle between the (red) rod $i$ and an arbitrary surrounding (black) rod $j$ is given by
\begin{equation}
\cos \theta_{ij} = \frac{\vec{v_{i}}\cdot\vec{v_{j}}}{|\vec{v_{i}}|\cdot|\vec{v_{j}}|}
\end{equation}
 
Figure \ref{fig:figure8} depicts two scenarios where the rods are mostly aligned with each other inside the cutoff $P_{2}\approx 1$ and when the rods are mostly randomly oriented $P_{2}\approx 0$.

\subsubsection{Global Order Parameter}
In a computer simulation the global order parameter may be computed in the following way
\begin{equation}
S = \frac{1}{N} \Bigg\langle \sum_{i}^{N} \Big(\frac{3}{2}\cos^{2}\theta_{i}\Big)-\frac{1}{2} \Bigg\rangle
\end{equation}
where $\theta_{i}$ is the angle of the $i$th backbone vector with the nematic director. The orientation of the nematic director is however already known from theory hence it is more useful to compute instead 
\begin{align}
S' &= \frac{1}{N} \Bigg\langle \sum_{i}^{N} (\frac{3}{2}(\boldsymbol{n}\cdot \boldsymbol{u}_{i})^{2}-\frac{1}{2} \Bigg\rangle\\
   &= \frac{1}{N}  \sum_{i}^{N}  \langle\boldsymbol{n}\cdot (\frac{3}{2}\boldsymbol{u}_{i}\boldsymbol{u}_{i} - \frac{1}{2}\boldsymbol{I})\cdot \boldsymbol{n}\rangle\\
   &= \frac{1}{N}  \sum_{i}^{N}  \langle\boldsymbol{n}\cdot \boldsymbol{Q}\cdot \boldsymbol{n}\rangle
\end{align}
where $\boldsymbol{n}$ is an arbitrary unit vector and $\boldsymbol{Q}_{i} = \frac{3}{2}\boldsymbol{u}_{i}\boldsymbol{u}_{i} - \frac{1}{2}\boldsymbol{I}$. The tensor order parameter is given by 
\begin{equation}
\langle \boldsymbol{Q} \rangle = \frac{1}{N} \sum_{i}^{N} \langle \boldsymbol{Q}_{i} \rangle 
\end{equation}

and is a traceless symmetric 2nd-rank tensor with three eigenvalues $\lambda_{+}$, $\lambda_{0}$ and $\lambda_{-}$. The nematic order parameter is defined as the largest positive eigenvalue of $\langle \boldsymbol{Q} \rangle$ and the true nematic director is its corresponding eigenvector \cite{eppenga1984monte}.

\subsection{\label{Appendix:MFT}Self-Consistent Theory}

The smectic coupling $\alpha$, is defined as 
\begin{eqnarray}
\alpha = 2\exp\Big[-\Big(\frac{\pi r_{0}}{d}\Big)^{2}\Big]
\label{SI_MFT_ALPHA}
\end{eqnarray}
where $r_{0}$ is the length of the rod-like LC molecule ($\approx 7.63\sigma$) and $d$ is the spacing between the smectic layers. The nematic and smectic order parameters $s$ and $\kappa$ are defined as
\begin{subequations}
\begin{eqnarray}
&s = \frac{1}{2} \langle 3\cos^{2}\theta-1\rangle \\*
&\kappa = \frac{1}{2}\Big\langle \cos\Big(\frac{2\pi z}{d}\Big)(3\cos^{2}\theta - 1) \Big\rangle
\end{eqnarray}
\end{subequations}
where $\theta$ represents the angle of an arbitrary rod-like polymer with the director and the angular brackets denote averages performed using the following translational-orientational distribution function,
\begin{eqnarray}
f(z,\cos\theta) = \frac{1}{4\pi \mathcal{Z}}\exp\Big[\frac{1}{2}m_{n}(3\cos^{2}\theta - 1)\Big]\\*
\nonumber
\times \exp\Big[\frac{1}{2}m_{s}\cos\Big(\frac{2\pi z}{d}\Big)(3\cos^{2}\theta - 1)\Big]
\end{eqnarray}
where $\mathcal{Z}$ is the partition function and $m_{n}$ and $m_{s}$ are the nematic and smectic mean-field parameters respectively which describe the potential field strength. In \cite{kyu1996phase} the partition function is defined as 
\begin{eqnarray}
\mathcal{Z}=\int^{1}_{-1}\int^{1}_{0}\exp\Bigg[\frac{1}{2}m_{n}\Big(3\cos^{2}\theta-1\Big)\Bigg] \label{SI_MFT_PARTITION} \\*
\nonumber
\times \exp{\Bigg[\frac{1}{2}m_{s}\cos\Big(\frac{2\pi z}{d}\Big) \Big(3\cos^{2}\theta-1\Big)\Bigg]} dz d\cos{\theta}
\end{eqnarray}
where $m_{n}$ and $m_{s}$ are dimensionless mean-field parameters which characterise the strength of the potential fields and correspond to the nematic and smectic phases respectively. Their order parameters $s$ and $\kappa$ may then be related to $\mathcal{Z}$ using the following relations as well as the entropy $\Sigma$
\begin{subequations}
\begin{eqnarray}
&s=\int^{1}_{-1}\int^{1}_{0}f(z,\cos{\theta})\frac{1}{2}\Big(3\cos^{2}\theta-1\Big) dz d\cos{\theta} \label{SI_MFT_NEMATIC} \\* 
\nonumber
&= \frac{1}{\mathcal{Z}}\frac{\partial\mathcal{Z}}{\partial m_{n}} \\*
&\kappa=\int^{1}_{-1}\int^{1}_{0}f(z,\cos{\theta})\frac{1}{2}\cos\Big(\frac{2\pi z}{d}\Big) \label{SI_MFT_SMECTIC}\\*
\nonumber
&\times \Big(3\cos^{2}\theta-1\Big) dz d\cos{\theta}=\frac{1}{\mathcal{Z}}\frac{\partial\mathcal{Z}}{\partial m_{s}} \\*
&\Sigma=-\int^{1}_{-1}\int^{1}_{0}f(z,\cos{\theta})\ln{[4\pi f(z,\cos{\theta})]}dzd\Omega \label{SI_MFT_ENTROPY} \\* 
\nonumber
&= \ln{\mathcal{Z}} - m_{n}s-m_{s}\kappa
\end{eqnarray}
\end{subequations}
where $\Omega$ denotes solid angle in Equation \ref{SI_MFT_ENTROPY}. The orientational order parameters $s$ and $\kappa$ are then evaluated by minimising the anisotropic portion of the free energy such that $\frac{\partial f_{aniso}}{\partial s}=0$ and $\frac{\partial f_{aniso}}{\partial \kappa}=0$ which results in the two coupled equations
\begin{subequations}
\begin{eqnarray}
&m_{s} = \alpha \nu(T) \kappa \phi \\*
\label{SI_SELFCON_1}
&m_{n} = \nu(T) s \phi
\label{SI_SELFCON_2}
\end{eqnarray}
\end{subequations}
which must be solved self-consistently. 

The numerical procedure for solving the mean-field theory begins by first solving Equation \ref{SI_MFT_PARTITION} for an initial guess of the mean-field parameters $m_{n}^{(0)}$ and $m_{s}^{(0)}$ at a given temperature $T$ and composition $\phi$. This is achieved by performing a 2d Simpsons rule integral and defining an approximate expression for the partition function
\begin{widetext}
\begin{eqnarray}
    \nonumber
    &\mathcal{Z}(m_{n},m_{s})\approx \frac{h_{1}h_{2}}{36} \sum_{i=0}^{N-1}\sum_{j=0}^{N-1} \Bigg\{
    f'\Big(ih_{1},jh_{2}\Big)+f'\Big((i+1)h_{1},jh_{2}\Big)+f'\Big(ih_{1},(j+1)h_{2}\Big)+f'\Big((i+1)h_{1},(j+1)h_{2}\Big) \\*
    \nonumber
    &+ 4\Big[f'\Big((i+1/2)h_{1},jh_{2}\Big)+f'\Big(ih_{1},(j+1/2)h_{2}\Big)2+f'\Big((i+1/2)h_{1},(j+1)h_{2}\Big)+f'\Big((i+1)h_{1},(j+1/2)h_{2}\Big)\Big] \\*
    & + 16f'\Big((i+1/2)h_{1},(j+1/2)h_{2}\Big)\Bigg\}
    \label{SI_MFT_SIMPSONS}
\end{eqnarray}
\end{widetext}
where $N$ is the number of Simpsons rule intervals, $f'(i,j)$ corresponds to the value of the expression inside the integral in Equation \ref{SI_MFT_PARTITION} and $h_{1}=2/N$ and $h_{2}=1/N$. Note the larger this number, the more computationally intensive the calculation becomes since this procedure must be repeated for every guess of $m_{n}$ and $m_{s}$ until a convergence condition is reached,  $N=100$ is sufficiently large for our purposes and gives sufficiently accurate statistics. For a given guess of $m_{n}$ and $m_{s}$ the expression in Equation \ref{SI_MFT_NEMATIC} is evaluated such that 
\begin{eqnarray}
    &s^{(0)}\approx\frac{1}{\mathcal{Z}(m_{n}^{(0)},m_{s}^{(0)})} \\*
    \nonumber
    &\times \frac{\mathcal{Z}(m_{n}^{(0)}+\delta m_{n},m_{s}^{(0)})-\mathcal{Z}(m_{n}^{(0)}-\delta m_{n},m_{s}^{(0)})}{2\delta m_{n}}
%\label{SI_MFT_SIMPSONS}
\end{eqnarray}
where $\delta m_{n}$ is some sufficiently small step, in our case $\delta m_{n}\approx 10^{-6}$ to give a value of the nematic order parameter $s^{(0)}$. This is then substituted into Equation \ref{SI_MFT_NEMATIC} to provide a new estimate of the mean-field parameter $m_{n}^{(1)}$ such that
\begin{eqnarray}
m_{n}^{(1)}=\nu(T) s^{(0)}\phi
\end{eqnarray}
At this point one can either reiterate the same procedure for the smectic ordering using the same initial guess for the mean-field parameters at step 0 ($m_{n}^{(0)}$ and $m_{s}^{(0)}$) or save iterations by using the new estimate $m_{n}^{(1)}$ in the next step, we choose the latter approach since it is more computationally efficient. Hence the smectic order parameter may be evaluated in much the same way
\begin{eqnarray}
    &\kappa^{(0)}\approx\frac{1}{\mathcal{Z}(m_{n}^{(1)},m_{s}^{(0)})} \\*
    \nonumber
    &\times \frac{\mathcal{Z}(m_{n}^{(1)},m_{s}^{(0)}+\delta m_{s})-\mathcal{Z}(m_{n}^{(1)},m_{s}^{(0)}-\delta m_{s})}{2\delta m_{s}}
%\label{SI_MFT_SIMPSONS}
\end{eqnarray}
and a new estimate for the smectic order parameter may be determined using 
\begin{eqnarray}
m_{s}^{(1)}=\alpha\nu(T) \kappa^{(0)}\phi
\end{eqnarray}
This procedure is then repeated using the new initial values for the mean-field parameters $m_{n}^{(1)}$ and $m_{s}^{(1)}$ until $|m_{n}^{(i)}-m_{n}^{(i+1)}|<=\xi$ and $|m_{s}^{(i)}-m_{s}^{(i+1)}|<=\xi$ where $\xi$ is some acceptable margin of error, in our case $\xi\approx10^{-6}$. Once this condition is met the free energy may be evaluated for a given value of $\phi$ and $T$. By repeating this procedure at fixed $T$ and solving for $m_{n}$ and $m_{s}$ at different values of $\phi$ one may evaluate the free energy $g(\phi)$ numerically for a given $T$. 

\begin{figure}[htb]
\begin{center}
\includegraphics[width=\columnwidth]{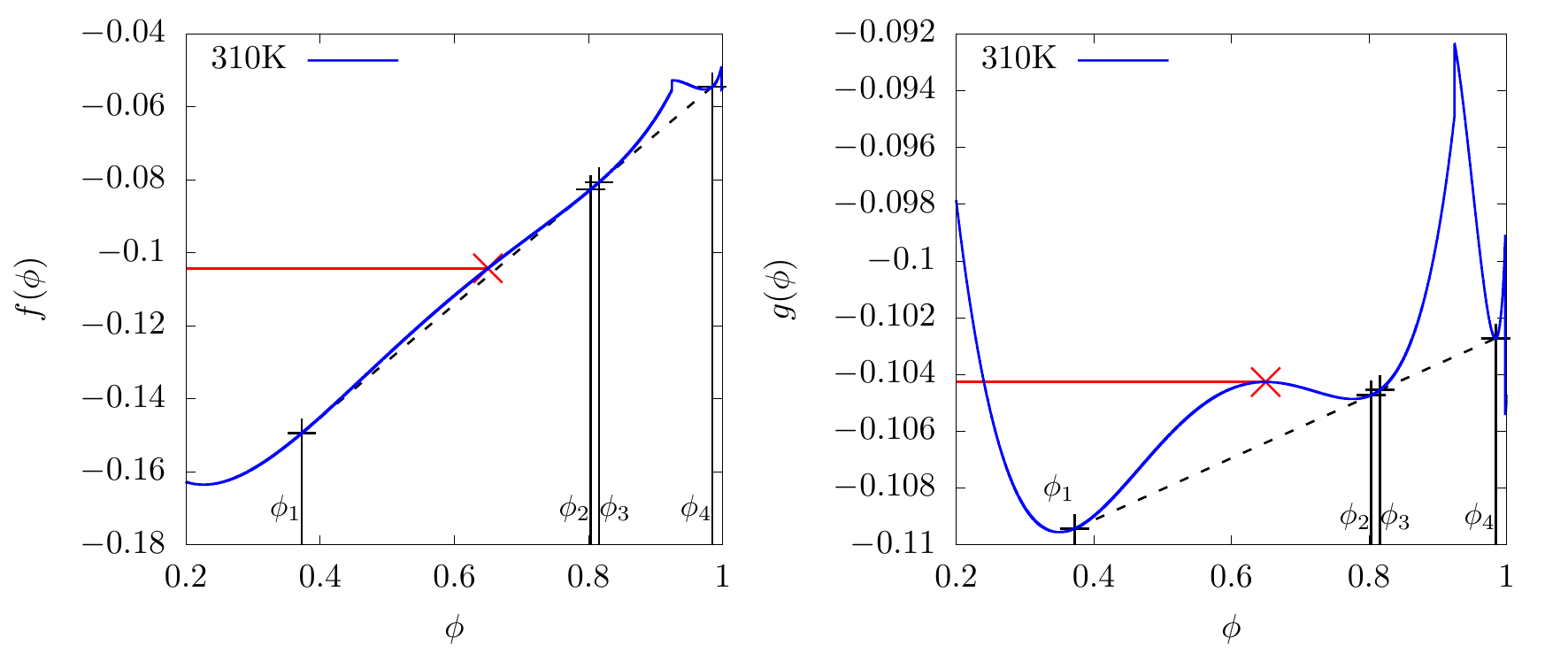} 
\caption{(left) Free energy $f(\phi)$, at 310K. (right) Transformed free energy $g(\phi)$. The red point in each figure denotes $\phi^{*}$ where $f(\phi^{*})=g(\phi^{*})$ and  $\frac{df(\phi)}{d\phi}\Bigr|_{\phi^{*}}\neq0$ and $\frac{dg(\phi)}{d\phi}\Bigr|_{\phi^{*}}=0$. Dashed lines indicate the common tangent solutions ($\chi=-1+772/T$, $T_{NI}=333K$, $\alpha=0.851$). }
\label{fig:figure9}
\end{center}
\end{figure}

\begin{figure}[htb]
\begin{center}
\includegraphics[width=0.998\columnwidth]{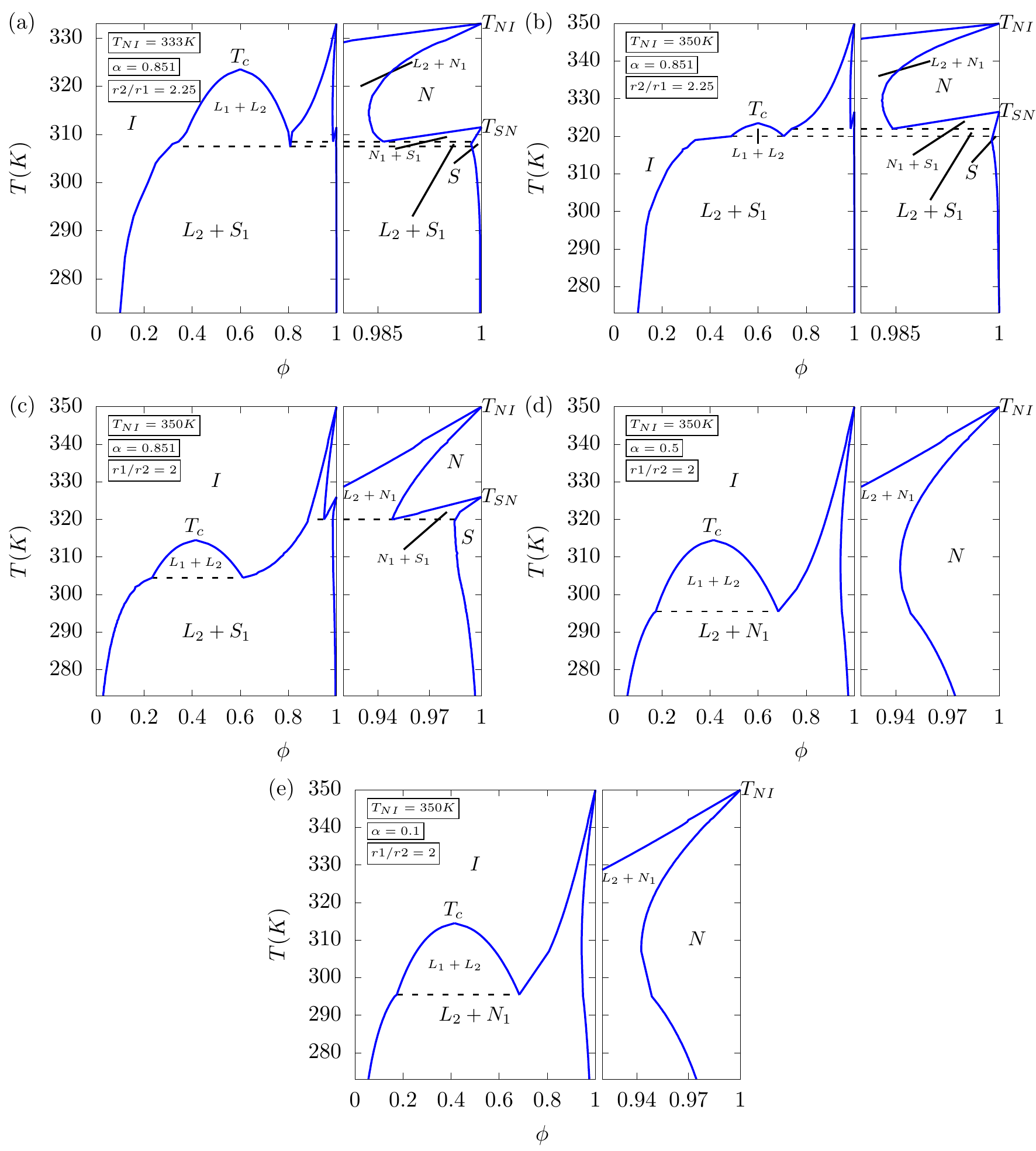} 
\caption{Temperature vs composition phase diagrams for a mixture of flexible polymers and rod-like smectic-A mesogens with (a) and (b) in the long polymer regime ($r_{1}/r_{2}=2.25$) and (c),(d) and (e) in the long mesogen limit ($r_{2}/r_{1}=2$). All parameters are given in the top left portion of the figures to which they correspond. In panel (a) the phase diagram originally presented in \cite{kyu1996phase} has been reproduced using identical parameters. In (b) $T_{NI}$ is raised by 17K from that originally presented in \cite{kyu1996phase} which effectively buries the $L_{1}+L_{2}$ coexistence region and extends the ``spout". In (c) the system switches into the long rod regime which shifts the $L_{1}+L_{2}$ region left and lowers $T_{c}$. In figures (d) and (e) the smectic interaction parameter $\alpha$ is much reduced and effectively switches off the smectic component of the free energy, reducing the model to a polymer nematic-liquid-crystal mixture.}
\label{fig:figure10}
\end{center}
\end{figure}

\begin{figure}[htb]
\begin{center}
\includegraphics[width=\columnwidth]{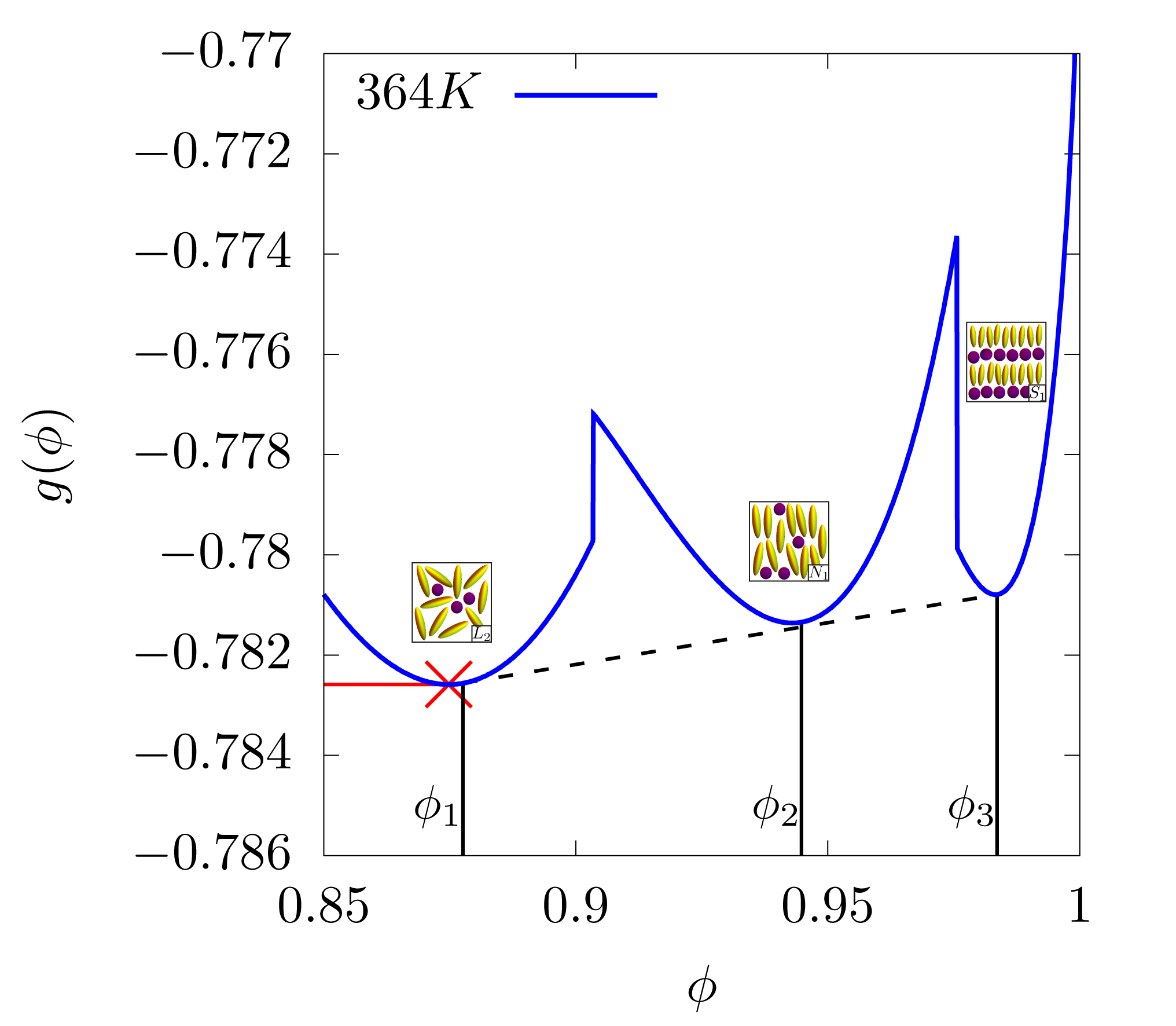} 
\caption{Transformed free energy $g(\phi)$, close to the triple point, at 364K. The red point in each figure denotes $\phi^{*}$ where $f(\phi^{*})=g(\phi^{*})$ and  $\frac{df(\phi)}{d\phi}\Bigr|_{\phi^{*}}\neq0$ and $\frac{dg(\phi)}{d\phi}\Bigr|_{\phi^{*}}=0$. Dashed lines indicate the common tangent solutions ($\chi=-1+772/T$, $T_{NI}=400K$, $\alpha=0.851$, $r_{1}/r_{2}$=2). }
\label{fig:figure11}
\end{center}
\end{figure}

A simple common tangent construction is then used to map out the phase diagram in the $T$-$\phi$ plane but this is in some cases a non-trivial exercise. Large linear terms dominate the free energy which if not dealt with appropriately impact the numerical precision of the gradient terms introducing a large error. This may be overcome by subtracting a linear gradient term from $f(\phi)$ such that
\begin{eqnarray}
g(\phi)=f(\phi)-\frac{df(\phi)}{d\phi}\Bigr|_{\substack{\phi^{*}}}(\phi-\phi^{*})
\label{SI_MFT_TRANSFORMATION}
\end{eqnarray}
where $\phi^{*}$ is some point along $f(\phi)$, $\frac{df(\phi)}{d\phi}$ is the first principles derivative evaluated at that point and $g(\phi)$ is the new free energy to be used in the common tangent construction. This procedure only improves the numerical precision of the common tangent construction and does not influence the position of the bracketing values $\phi_{1}$ and $\phi_{2}$. At these points the chemical potential $\mu(\phi)=\frac{df(\phi)}{d\phi}$ and osmotic pressure $\Pi(\phi)=\phi\frac{df(\phi)}{d\phi}-f(\phi)$ are equal, such that the following equilibrium conditions are satisfied.
\begin{subequations}
\begin{eqnarray}
\label{SI_MFT_EQUIL1}
&\mu(\phi_{1})=\mu(\phi_{2}) \\*
\label{SI_MFT_EQUIL2}
&\Pi(\phi_{1})=\Pi(\phi_{2})
\end{eqnarray}
\end{subequations}
It may be proven that this condition holds before and after applying the transformation in Equation \ref{SI_MFT_TRANSFORMATION} to demonstrate that $\phi_{1}$ and $\phi_{2}$ are invariant. Under the transformation, Equations \ref{SI_MFT_EQUIL1} and \ref{SI_MFT_EQUIL2} may be rewritten as 
\begin{subequations}
\begin{eqnarray}
&\mu'(\phi) = \mu(\phi)-\frac{df(\phi)}{d\phi}\Bigr|_{\substack{\phi^{*}}} = \mu(\phi)-\mu(\phi^{*}) \\*
&\Pi'(\phi)=\underbrace{\phi\frac{df(\phi)}{d\phi}-f(\phi)}_{\Pi(\phi)} - \underbrace{\phi^{*}\frac{df(\phi)}{d\phi}\Bigr|_{\substack{\phi^{*}}}}_{\Pi^{*}} \\*
\nonumber
& = \Pi(\phi) - \Pi(\phi^{*})
\end{eqnarray}
\end{subequations}
where $\mu'(\phi)$ and $\Pi'(\phi)$ are the chemical potential and osmotic pressure after the transformation. Thus the points $\phi_{1}$ and $\phi_{2}$ are invariant under the transformation and 
\begin{subequations}
\begin{eqnarray}
&\mu'(\phi_{1})=\mu'(\phi_{2}) \\*
%\label{SI_MFT_EQUIL1}
&\Pi'(\phi_{1})=\Pi'(\phi_{2})
%\label{SI_MFT_EQUIL2}
\end{eqnarray}
\end{subequations}

\subsection{\label{Appendix:MFT}Characteristics of the Mean-Field Phase Diagram }
The mean-field phase diagrams, in the $T-\phi$ plane, resulting from following the above computational details are presented in Figure \ref{fig:figure10} as we systematically vary the parameters in order to gradually transition from the physical situation presented in \cite{kyu1996phase} to a set of parameter values which is close to that which is appropriate for describing our CGMD simulations. Figure \ref{fig:figure10} (panel (a))shows the phase diagram for short mesogens and longer polymers and it has the ``classic'' tea-pot shape with well-separated lid (terminating at the critical point with $T_{c} \sim$ 320 K) and the double spout regions, with the upper spout characterising the isotropic to nematic transition close to $\phi = 1$ and at $T = T_{NI}$ and the lower spout characterising the transition from nematic to the smectic state close to $\phi = 1$ and at $T = T_{SN}$. The two dashed, horizontal lines denote the two closely located triple points, a temperature at which the three phases coexist. In going from panel (a) to (b) the effect of the increase of the temperature $T_{NI}$ on the shape of the phase diagram has been studied. One observes that the upper spout thus originates from the new higher value of $T_{NI} = 350 K$ and as a result the $L_{2}-N_{1}$ coexistence region starts encroaching into lower $\phi$ values and as a result of this the lid of the teapot gets buried into the encroaching $L_{2}-N_{1}$ coexistence region. In panel (c) the mesogens have been made longer than the polymers, in accordance with our CGMD simulations and we observe that the critical region, signified by the lid of the tea-pot has been pushed to lower $\phi$ values, compared to the situation in panel (a), and as a result the effect of anisotropic phases starts occurring from lower $\phi$ values. The second spout which occurs due to the occurrence of the smectic phase is controlled by the value of the parameter $\alpha$. In going from panel (c) to (d) the value of $\alpha$ has been reduced, leading to the complete disappearance of the smectic phase from the phase diagram. A similar trend of the missing smectic phase, is shown in panel (e) upon a further reduction of the parameter $\alpha$.     

\bibliography{fall_etal_bibliography.bib}

\end{document}